\documentclass[3p,12pt]{elsarticle}
\usepackage{threeparttable,booktabs,tabularx}
\usepackage[fleqn]{amsmath}
\usepackage{cases}
\usepackage{multirow}
\usepackage{xcolor}
\usepackage[normalem]{ulem}
\usepackage{tabularx}
\usepackage{siunitx}
\usepackage{comment}
\usepackage{algorithm}
\usepackage{algpseudocode}
\usepackage{algorithmicx}
\usepackage{grffile}
\usepackage{rotating}
\usepackage{array}
\usepackage{lineno}
\usepackage{hyperref}
\usepackage{makecell}
\usepackage{graphicx,enumerate}
\usepackage{amssymb}
\usepackage{bm,amsmath}
\usepackage{subfigure}
\usepackage{caption,color}
\usepackage{threeparttable}
\usepackage{subeqnarray}
\usepackage{multirow}
\usepackage{lscape}
\usepackage{rotating}
\usepackage{graphics}
\floatname{algorithm}{Algorithm}
\usepackage{epstopdf}
\usepackage{adjustbox}
\journal{}
\usepackage{bm}

\biboptions{numbers,sort&compress} % 压制引用，[1,2,3]--> [1-3]

\usepackage{algorithm}      % 提供 algorithm 环境
\usepackage{algpseudocode}  % 提供 \For, \If, \State 等伪代码命令 (需要 algorithmicx 包)
\usepackage{algorithmicx}   % algorithmicx 包本身 (algpseudocode 依赖于它)

\begin{document}
	
\begin{frontmatter}

\title{Accelerating the Monte Carlo simulation of the Enskog equation for multiscale dense gas flows}
%Multiscale simulation of hard-sphere dense gas flow via DIG method
\author{Bin Hu}
\author{Liyan Luo}  
\author{Lei Wu\corref{mycorrespondingauthor1}}
\ead{wul@sustech.edu.cn}
\cortext[mycorrespondingauthor1]{Corresponding author}
\address{Department of Mechanics and Aerospace Engineering,
Southern University of Science and Technology, 518055 Shenzhen, China}

\begin{abstract}
A general synthetic iterative scheme is proposed to solve the Enskog equation within a Monte Carlo framework. The method demonstrates rapid convergence by reducing intermediate Monte Carlo evolution and preserves the asymptotic-preserving property, enabling spatial cell sizes much larger than the mean free path in near-continuum flows. This is realized through mesoscopic–macroscopic two-way coupling: the mesoscopic Monte Carlo simulation provides high-order constitutive relations to close the moment (synthetic) equation, while the macroscopic synthetic equation, once solved toward steady state, directs the evolution of simulation particles in the Monte Carlo method. 
%To further reduce sampling noise, an exponentially weighted moving time average is applied, and the synthetic equation is updated every 100 Monte Carlo time steps to balance accuracy and efficiency. 
The accuracy of the proposed general synthetic iterative scheme is verified through one-dimensional normal shock wave and planar Fourier heat transfer problems, while its fast-converging and asymptotic-preserving properties are demonstrated in the force-driven Poiseuille flow and two-dimensional hypersonic cylinder flow and low-speed porous media flow, where the simulation time is reduced by several orders of magnitude in near-continuum flows. With the proposed method, a brief analysis is conducted on the role of the adsorption layer in porous media flow, mimicking shale gas extraction.
\end{abstract}


\end{frontmatter}

\section{Introduction}\label{sec:1}
The non-equilibrium dynamics of dense gases or at gas–liquid interfaces have attracted significant attention in recent years~\cite{Visible-light2024,evaporative_JFM_2024,Henning_Frezzotti_2022,Graru_Sharipov_ijhmt}. 
This scenario arises in various applications, including high-pressure shock tubes~\cite{petersen2001nonideal}, shale gas extraction~\cite{wu2016non, PHS-MD}, gas–liquid mixing in high-pressure injection systems~\cite{dahms2015}, and evaporation/condensation processes~\cite{frezzotti2005mean, kon2014method}. 
These non-equilibrium (rarefied gas) flows are primarily characterized by the Knudsen number (Kn), defined as the ratio of the molecular mean free path to a characteristic flow length. The Navier–Stokes (NS) equations can adequately predict slightly rarefied gas flows by incorporating velocity slip and temperature jump boundary conditions. However, when Kn becomes appreciable, these equations deviate significantly from experimental observations because the linear constitutive relations (e.g., the Newton law of viscosity and Fourier law of heat conduction) lose accuracy~\cite{Ewart2007JFM}. The Boltzmann equation, which directly models molecular streaming and collisions, captures the gas dynamics across the full range of rarefaction~\cite{CE1970}. However, the Boltzmann equation becomes inaccurate for non-ideal gases, particularly when the molecular size becomes comparable to the mean free path. 
The Enskog equation~\cite{enskog_1922}, which extends the Boltzmann equation by incorporating nonlocal collisions, models the dense non-ideal gases composed of hard-sphere molecules~\cite{CE1970, ferziger_kaper_1972}. Further, with the mean field force, the Enskog-Vlasov equation is used to model the liquid-gaseous systems~\cite{evaporative_JFM_2024, Henning_Frezzotti_2022, Graru_Sharipov_ijhmt}.

As with the Boltzmann equation, two primary methods are employed for solving the Enskog equation numerically: the deterministic and stochastic methods. The discrete velocity method, where the velocity distribution function is discretized in both spatial and velocity domains, enables the deterministic numerical simulation with mature techniques in computational fluid dynamics. For example, Wu \textit{et al} proposed a fast spectral method for solving the Enskog collision integral~\cite{wu-dense-FSM2015}; together with the conventional iterative scheme, they reported novel dynamics of highly-confined Poiseuille flow of dense gas~\cite{wu2016non}. In contrast, the stochastic methods model gas flow by representing the gas as a collection of simulation particles whose free streaming and collisions are decoupled. For instance, Frezzotti and Sgarra~\cite{frezzotti1993shockwave} introduced a Monte Carlo quadrature method for the Enskog equation based on Nanbu’s scheme~\cite{Nanbu_scheme}. Later, Frezzotti~\cite{frezzottiAlgorithm1997} proposed a particle-based method derived from the direct simulation Monte Carlo (DSMC) approach, which accurately conserves momentum and energy. This method was validated by comparing molecular dynamics results for heat flow in dense hard-sphere gases. Alexander\textit{et al} developed the consistent Boltzmann algorithm~\cite{CBA1995}, which increases the collision rates by introducing an additional displacement during convection. Although this modification is based on intuitive reasoning rather than a formal derivation, the resulting solutions remain valid across a wide range of fluid densities. It preserves thermodynamic and transport properties in low-density regimes and aligns with Enskog-based models in high-density conditions.
Montanero and Santos~\cite{ESMC1} introduced the Enskog simulation Monte Carlo (ESMC) method, which improves the collision probability by selecting colliding pairs from cells separated by a distance equal to the particle diameter, thereby accounting for spatial correlations. This method accurately reproduces the transport properties of dense gases but, like Nanbu’s method~\cite{Nanbu_scheme}, only conserves momentum and energy in a statistical sense. They later enhanced the ESMC method by incorporating Bird’s no-time-counter scheme, enabling complete conservation of momentum and energy~\cite{ESMC2}.

Two strategies are adopted to improve the simulation efficiency for dense gas flows. 
The first is to simplify the Enskog collision operator. For instance, Luo~\cite{Luo1998} developed an isothermal non-ideal gas lattice Boltzmann model by simplifying the Enskog collision integral using the Chapman-Cowling method~\cite{CE1970}. However, this model only applies to isothermal flows and is constrained by the characteristics of the lattice Boltzmann method, which is only suitable for near-continuum and incompressible fluid flow. Subsequently, Guo \textit{et al} proposed a simplified kinetic model for strongly inhomogeneous flows~\cite{zhaoliGuo2005,zhaoliGuo2006}. This model describes the density inhomogeneity on the basis of the concepts inspired by density-functional theory and the Fischer and Methfessel model for inhomogeneous fluids~\cite{Fischer_Methfessel_1980}. %However, this approach fails to account adequately for thermal and non-equilibrium effects, leading to its validity only for equilibrium and isothermal flows. 
Wang \textit{et al}.~\cite{wang2020} introduced the non-isothermal Shakhov-Enskog model for non-equilibrium dense gas flow, validated by comparisons with the fast spectral method~\cite{wu-dense-FSM2015}. This model focuses on capturing correct shear viscosity but neglects accurate recovery of thermal and bulk viscosity. 
%Busuioc~\cite{Busuioc_2023} further advanced this work by developing the Shakhov-Enskog-LBM model, which demonstrated good computational efficiency and accuracy when contrasted with particle-based Enskog equation solutions. Nevertheless, this study was limited to non-confined gases with inconsistent transport coefficients. 
To obtain transport coefficients consistent with the Enskog equation, Su \textit{et al} refined the Shakhov-Enskog model by incorporating density oscillation effects caused by gas-solid wall interactions, achieving results consistent with the Enskog equation and molecular dynamics simulations~\cite{su2023kinetic}. Alternatively, Sadr and Gorji proposed the Fokker-Planck model capable of simulating nonequilibrium dense gas flows~\cite{SadrGorji2017}.  

The second strategy is to develop multiscale methods. This approach is motivated by the fact that nearly all deterministic and stochastic models treat molecular streaming and collisions as separate processes; as a result, they are constrained by the requirement that the cell size and time step must remain smaller than the mean free path and the mean collision time, respectively. This does not pose a problem when the Knudsen number is large; however, when the Knudsen number is small, not only does significant numerical dissipation arise, but also the solution converges extremely slowly. For instance, in the one-dimensional Poiseuille flow with $\text{Kn}\sim10^{-3}$, the conventional iterative scheme requires approximately one million steps to converge. Worse still, the resulting solution is highly susceptible to numerical dissipation if the spatial grid is not sufficiently refined~\cite{wang2018comparative}.

Consequently, developing a computationally efficient and accurate solution strategy for the Enskog equation is imperative. Recently, the general synthetic iterative scheme (GSIS) has been proposed to solve the Shakhov-Enskog equation efficiently~\cite{gsis_Dense},  which alternately solves the  mesoscopic kinetic equation and macroscopic synthetic equation, not only demonstrating minimal numerical dissipation with coarse spatial grids, but also exhibiting fast convergence across all flow regimes. 

In this paper, we are going to apply this idea to enhancing the ESMC simulations. Note this idea has been recently proposed to boost the DSMC method for dilute gas, i.e.,  the direct intermittent GSIS (DIG) method~\cite{DIG, DIG_poly}, a simple numerical framework possessing both asymptotic-preserving and fast-convergence properties for efficient and accurate stochastic simulations of rarefied gas dynamics. In DIG, the solution of macroscopic synthetic equations is intermittently applied to the DSMC simulation, typically at intervals of 100 time steps, through a simple linear transformation of particle velocities and an addition and deletion of particles. 
With minimal modifications to the standard DSMC framework, DIG not only eliminates the restrictions on spatial cell size and time step, but also achieves rapid convergence to steady-state solutions, thereby reducing computational time in near-continuum flows by several orders of magnitude.

The remainder of the paper is organized as follows. Section \ref{sec:2} introduces the Enskog equation for dense gas and the ESMC method. Section \ref{sec:4} designs the high-order constitutive relations in the macroscopic synthetic equation, and details the implementation of the DIG algorithm to accelerate the ESMC simulations. Section \ref{sec:5} validates the DIG method through simulations of the one-dimensional normal shock wave and planar Fourier/Poiseuille flow, the two-dimensional hypersonic flow past a cylinder, and the dense flow through porous media. Finally, Section \ref{sec:6} provides concluding remarks and an outlook.

\section{The Enskog equation and particle method}\label{sec:2}

We present the Enskog equation for a dense monatomic hard-sphere gas, the associated macroscopic moment equations, and the ESMC algorithm used to solve them.

% The DIG method, like GSIS, employs an alternating application of mesoscopic and macroscopic solvers.

\subsection{Kinetics of dense gas}

Enskog’s theory rests on two key premises~\cite{enskog_1922}. First, colliding molecules are not taken to occupy the same point; instead, their centers are separated by exactly one hard-sphere diameter at contact. Second, the finite volume occupied by gas molecules reduces the space for motion, thereby increasing the collision frequency. In the absence of external forces, the Enskog equation for hard-sphere molecules reads:
\begin{equation}\label{EE}
\begin{aligned}
    \frac{\partial f}{\partial t}+\bm v \cdot \frac{\partial f}{\partial \bm x} 
    +\bm a \cdot \frac{\partial f}{\partial \bm v} = \sigma^2 \int\{\chi&\left[\bm{x}+{\sigma}\bm{k}|n\right] f\left(\bm{x}, \bm v^{\prime}, t\right) f(\bm{x}+\sigma \bm{k},\bm v_1^{\prime},t) \\
        -\chi&\left[\bm{x}-{\sigma}\bm{k}|n\right] f(\bm{x},\bm v,t) f(\bm{x}-\sigma \bm{k},\bm{v}_1,t)\} \varTheta(\bm{g} \cdot \bm{k})\bm{g} \cdot \bm{k} {d} \bm{k} {d} \bm v_1.
\end{aligned}
\end{equation}
Here, $f(\bm{x}, \bm v, t)$ represents the velocity distribution function, where $t$ is the time, $\bm{x}$ is the spatial coordinates, and $\bm v$ is the molecular velocity space; $\bm a$ is the external acceleration; $\sigma$ denotes the effective molecular diameter, which is determined by the viscosity; $\bm g=\bm v_1-\bm v$ is the relative velocity of two colliding molecules, where $\bm v$ and $\bm v_1$ are the molecular velocities before the collision, and $\bm k$ is a unit vector specifying their relative position at the time of impact. The post-collision velocities $\bm v^{\prime}$ and $\bm v_1^{\prime}$ are related to the precollision velocities through $\bm v^{\prime}=\bm v-\bm{k}(\bm{g} \cdot \bm{k})$ and $\bm v_1^{\prime}=\bm v_1+\bm{k}(\bm{g} \cdot \bm{k})$. $\varTheta(x)$ is the Heaviside function.

In the standard Enskog theory, the two-point pair-correlation function $\chi\left[\bm{x},\bm x \pm \sigma \bm k|n\right]$ measures the enhanced collision frequency due to the volume exclusion effect; it is evaluated based on the local density $n$ at the contact point $\bm x \pm \sigma\bm k/2$ of the collision pair~\cite{modifiedEnskog1973}. The function provides the probability of finding molecules at positions $\bm x$ and $\bm x \pm \sigma \bm k$ within a density field at equilibrium and can be defined as follows:
\begin{equation}\label{PCF}
    \begin{aligned}[b]
    \chi(\bm x \pm \sigma \bm k|n) = \chi\left[n(\bm x \pm \frac{\sigma}{2} \bm k)\right], \quad\quad
    \chi(n) = \frac{1-0.125bn}{(1-0.25bn)^3},\quad\quad b=\frac{2\pi\sigma^3}{3},
    \end{aligned}
\end{equation}
The expression ensures that the equation of state for the hard-sphere fluid aligns with the Carnahan-Starling formula~\cite{Carnahanstarling}:
\begin{equation}
  P = nk_BT\left( 1+bn\chi \right).
\end{equation}

The mean free path and the Knudsen number of the dense gas are defined as:
\begin{equation}\label{mfp}
    \begin{aligned}
        \lambda_0 = \frac{1}{\sqrt{2}\pi n \sigma^2 \chi(n)}, 
        \quad 
        \text{Kn}=\frac{{\lambda_0}}{L} = \frac{1}{\sqrt{2}\pi n \sigma^2 \chi(n) L}.
    \end{aligned}
\end{equation}
Furthermore, in order to describe the degree of denseness of the gas, the Enskog number En is introduced as the ratio of the molecular diameter $\sigma$ to the mean free path $\lambda$:
\begin{equation}\label{dimensionless_parameter}
        \begin{aligned}[b]
           \text{En} = \frac{\sigma}{\lambda_0} = \frac{3\sqrt{2}}{2}bn \chi(n).
        \end{aligned}
\end{equation}

\subsection{Macroscopic equation}

When $f(t, \bm{x}, \bm{v})$ is solved,  macroscopic quantities can be obtained by taking its moments. Specifically, the density $\rho$, velocity $\bm{u}$, temperature $T$, the kinetic stress tensor $\bm{P}_k$, and heat flux $\bm{q}_k$ can be calculated as:
\begin{equation}\label{macro_distribution}
\left[\rho,\; \rho\bm{u},\; \frac{3}{2}\rho RT,\; \bm{P}_k,\; \bm{q}_k\right] = 
\int \left[1,\; \bm{v},\; \frac{m}{2}c^2,\; m\bm{c}\bm{c},\; \frac{m}{2}c^2\bm{c}\right] f(t,\bm{x},\bm{v})  d\bm{v}
\end{equation}
where $m$ denotes the molecular mass, $R = k_B/m$ is the Boltzmann constant, and $\bm{c} = \bm{v} - \bm{u}$ represents the peculiar velocity. The kinetic stress tensor $\bm{P}^k$ and heat flux $\bm{q}^k$ result from the free motions of molecular momentum and energy. 

On multiplying Eq.~\eqref{EE} by $\phi = (m, m\bm v, \frac{1}{2}m\bm v^2)$ and integrating the resulting equations over the whole molecular velocity space, the macroscopic synthetic equations for the conservation of mass, momentum, and energy are eventually obtained:
\begin{equation}\label{macro}
    \begin{aligned}
    \frac{\partial \rho}{\partial t} +{\nabla} \cdot(\rho \bm{u}) & =0, \\
    \frac{\partial (\rho\bm{u})}{\partial t}+{\nabla}\cdot(\rho \bm{u} \bm{u}+\bm P_k + \bm P_{c}) & =0, \\
    \frac{\partial (\rho e)}{\partial t}+ {\nabla} \cdot(\rho e \bm{u}+\bm{\bm P_k} \cdot \bm{u}+\bm P_{c}\cdot \bm{u}+\bm{q}_k +\bm q_{c}) & =0,
    \end{aligned}
\end{equation}
where $\rho=nm$ is the mass density of the gas, and $e=\frac{3 k_B}{2 m} T+\frac{1}{2} u^2$ is the total energy per unit mass of gas. 

It should be emphasized that, unlike in dilute gases, $m\bm v$ and $\tfrac{1}{2}m\bm v^2$ are not collisional invariants of the Enskog collision operator. Consequently, the transfer of momentum and energy during collisions plays an important role in determining the dynamics of dense gases. These effects are incorporated into the potential parts of the stress and heat flux in Eq.~\eqref{macro}, which are defined as follows~\cite{Kremer2010book}:
\begin{equation}\label{potential_pq}
    \begin{aligned}
    \{{\bm {P}}_{c},{\bm {q}}_{c}\} &= \frac{\sigma^2}{2}\iiint d \boldsymbol{v} d \boldsymbol{v}_1 d \boldsymbol{k}\int_{0}^{\sigma}d \alpha~\chi\bigg[n\bigg(\bm{x}+\alpha{\bm{k}}-{\frac{\sigma}{2}}{\bm {k}}\bigg)\bigg]  (\psi^{\prime}-\psi)\\
    & \times f(\boldsymbol{x}+\alpha \boldsymbol{k}, \boldsymbol{v},t) f\left(\boldsymbol{x}+\alpha \boldsymbol{k}-\sigma \boldsymbol{k}, \boldsymbol{v}_1,t\right)(\bm{g} \cdot \bm{k})\varTheta(\bm{g} \cdot \bm{k}),
    \end{aligned}
\end{equation}
where $\psi = \{m\bm c, ~\frac{1}{2}m\bm c^2\}$, $\psi'=\{m\bm c^{\prime}, ~\frac{1}{2}m\bm c^{\prime 2}\}$, and $\alpha$ is a dummy variable.

\subsection{The ESMC method}

Similar to the DSMC method for the Boltzmann equation, ESMC employs a representative ensemble of simulation particles to mimic the streaming and collision of dense gas molecules described by the Enskog equation~\cite{ESMC1}. 
In ESMC, each simulation particle represents $N_{{eff}}$ real gas molecules, each carries information about its location $\bm x$, pre-collision velocity $\bm{v}$, post-collision velocity $\bm{v}^{\prime}$, and a random movement direction unit vector $\bm {k}$. These particles transport through the simulation domain, which is discretized into computational cells. The local macroscopic properties, including the density $\rho$, flow velocity $\bm{u}$, temperatures $T$, the kinetic/potential stress tensors, and the kinetic/potential heat fluxes, are sampled  within each cell of volume $V_{{cell}}$~\cite{ESMC1,ESMC2}:
\begin{equation}\label{ESMC_macro}
    \begin{aligned}
      &\rho=\frac{N_{eff}}{V_{cell}}N_{p},\quad   u_{i}=\frac{1}{N_{p}}\sum_{p=1}^{N_{p}}v_{i,p}^{},\quad   T=\frac{1}{3N_{p}}\sum_{p=1}^{N_{p}}\left|\bm v_{p}-\bm u\right|^{2},\\
      &\sigma_{k,ij}=\frac{N_{eff}}{V_{cell}}\sum_{p=1}^{N_{p}}\left[\left(v_{i,p}-u_{i}\right)\left(v_{j,p}-u_{j}\right)-\frac{\delta_{ij}}{3}|\bm v_p-\bm u|^2\right],\\
      &q_{k,i}=\frac{N_{eff}}{2V_{cell}}\sum_{p=1}^{N_{p}}\left(v_{i,p}-u_{i}\right)\left|\bm v_{p}-\bm u\right|^{2},\\
      &\sigma_{c,ij}=\frac{N_{eff}\sigma}{V_{cell}\Delta t}\sum_{p=1}^{N_{p}}\left[\left(v_{i,p}-v^{\prime}_{i,p}\right)k_{j,p}-\frac{\delta_{ij}}{3}(\bm v_p-\bm v_p^\prime)\cdot\bm k_p\right],\\
      &q_{c,i}=\frac{N_{eff}\sigma}{2V_{cell}\Delta t}\sum_{p=1}^{N_{p}}\left[\left(\bm v_{p}^2-\bm v^{\prime2}_{p}\right)k_{i,p}-2(\bm k_p\cdot \bm u)(v_{i,p}- v_{i,p}^\prime)\right],
    \end{aligned}
\end{equation}
where $N_p$ denotes the number of simulation particles in the cell, the subscripts $i, j=x,y,z$ represent the spatial direction, and $\delta_{ij}$ is the Kronecker delta function.

The primary distinction between ESMC and DSMC lies in their treatment of the collision process. 
The present work employs the ESMC algorithm proposed in Ref.~\cite{ESMC2}, which extends Bird's no-time-counter collision scheme~\cite{bird-1994} to the Enskog framework. %By modifying the collision frequency to account for potential collisions between particles residing in different computational cells, the improved algorithm rigorously enforces exact conservation laws. 
The collision probability for a pair of particles $i$ in cell $I$ and $j$ in cell $J$ is given by:
\begin{equation}
\omega_{ij}={4\pi\sigma^2\varTheta(\bm g_{ij} \cdot \bm k_i)(\bm  g_{ij}\cdot \bm k_i)\chi\left(n_{\text {mid}}\right)n_j\Delta t},
\end{equation}
where ${\bm g_{ij}} =\bm v_i- \bm v_j$ is the relative velocity of the collision pair, $n_j$ is the number density of cell $J$, $n_{\text{mid}}$ is the number density evaluated at a midpoint of collision, and $\bm k_i$ is the random unit direction vector of particle $i$. When particle $i$ collides with particle $j$, the velocities of the collision pair evolve as follows:
\begin{equation}
    \begin{aligned} \label{post_velocity}
    &\bm v_i^{\prime}=\bm v_i-(\bm k_i\cdot\bm g_{ij})\bm k_i, \quad \bm v_j^{\prime}=\bm v_j+(\bm k_i\cdot\bm g_{ij})\bm k_i,\\
    &\bm k_i = [\cos (\theta), ~\sin (\theta) \cos (\phi), ~\sin (\theta) \sin (\phi)],
    \end{aligned}
\end{equation}
where the direction angles $\theta= \arccos(2r_1-1)$ and $\phi =2\pi r_2$ are sampled in spherical coordinates and $r_1, r_2$ are random numbers sampled from a continuous uniform distribution in the interval [0, 1].

For elastic collisions within the timestep $\Delta t$, the initial number of candidate collisions for cell $I$ containing $N_I$ particles is $N^{coll}_{I} = \frac{1}{2} N_I \omega_{\text{max}}^I$. The initial upper bound for the collision probability is estimated using the molecular thermal velocities during initialization: $\omega_{\text{max}}^I = 4 \pi \sigma^2 n_I \Theta_I \chi\left(n_I\right) \Delta t$, where $\Theta_I = 10 \sqrt{{k T_0^I}/{m}}$. During each timestep, if $\omega_{ij} > \omega_{\text{max}}^I$, $\omega_{\text{max}}^I$ is updated to $\omega_{ij}$.

However, simply iterating over cells and executing Enskog collisions, analogous to typical DSMC implementations, may introduce bias due to cross-cell collisions and the prioritization of particles in cells processed earlier in ESMC. Therefore, we utilize an unbiased particle selection algorithm for the ESMC method~\cite{Sadr_Pfeiffer_Gorji_2021}.  It should be noted that if the point $\bm x_i + \sigma \bm k$ exceeds the domain boundary during the collision process, this indicates that the distance from the particle to the boundary is less than $\sigma$. In such cases, no collision occurs.

\section{DIG for dense gas}\label{sec:4}%%%%%%%%%%%%%%%%%%%%

As will be observed in the next section, the ESMC works well when the Knudsen number is large. However, when the Knudsen number is small, a large number of spatial cells is required to reduce numerical dissipation, and an enormous number of time steps is needed to reach the steady-state solution. To address this, the core idea of GSIS is developed to promote fast convergence and ensure asymptotic preserving properties.

In this section, we first derive the macroscopic synthetic equations from the Enskog equation, with a detailed explanation of the construction of constitutive relations that remain valid across the entire range of gas rarefaction. Subsequently, the DIG algorithm is proposed, where the synthetic equations are intermittently coupled with the ESMC to enhance computational accuracy and efficiency.

\subsection{Macroscopic synthetic equations}
To inherit the fast convergence and asymptotic-preserving properties of the GSIS, the macroscopic synthetic equations must be solved in conjunction with the ESMC simulation. We note that Eq.~\eqref{macro} is not closed unless constitutive relations for the stress tensor and heat flux are given. In the continuum flow limit, the Chapman-Enskog expansion of the kinetic equation gives the NS constitutive relations as~\cite{CE1970}:
\begin{equation}\label{NS-relations}
\begin{aligned}
&\bm P^{\text{NS}} =\bm P_{k}^{\text{NS}}+\bm P_{c}^{\text{NS}} = (P-\zeta_*{\nabla}\cdot\bm{u})\bm I+\bm \sigma_{k}^{\text{NS}}+\bm \sigma_{c}^{\text{NS}} ,\\
&\bm\sigma_{k}^{\text{NS}}=-\mu_k\left(\nabla \bm u+\nabla \bm u^\mathrm{T}-\frac{2}{3}{\nabla}\cdot\bm{u}\bm I\right),\\
&\bm \sigma_{c}^{\text{NS}}=-\mu_c\left(\nabla \bm u+\nabla \bm u^\mathrm{T}-\frac{2}{3}{\nabla}\cdot\bm{u}\bm I\right),\\
&\bm q_k^{\text{NS}} = -\kappa_k\nabla T, \quad
\bm q_c^{ \text{NS} } =-\kappa_c{\nabla}T,\\
\end{aligned}
\end{equation}
where $\zeta_{*}=\mu_{*}\chi(bn)^2$ is the bulk viscosity due to the non-local collision, $\bm I$ is the unit matrix, and the shear viscosity and thermal conductivity of the dense gas are given by~\cite{ESMC1}:
\begin{equation}\label{mu_kap}
\begin{aligned}
&\mu_k = \frac{\mu_{*}}{\chi}\left(1+\frac{2}{5}bn\chi\right),\quad \mu_c=\frac{\mu_{*}}{\chi}\left(1+\frac{2}{5}b n\chi\right)\frac{2}{5}bn\chi+\frac{3}{5}\zeta_{*},\\
&\kappa_k=\frac{\kappa_{*}}{\chi}\left(1+\frac{3}{5}b n\chi\right),\quad\kappa_c=\frac{\kappa_{*}}{\chi}\left(1+\frac{3}{5}bn\chi\right)\frac{3}{5}bn\chi+\frac{3}{2}\frac{k_{B}}{m}\zeta_{*},
\end{aligned}
\end{equation}
with 
$\mu_{*}$ and $\kappa_{*}$ being the shear viscosity and thermal conductivity in the dilute gas limit: 
\begin{equation}
\begin{aligned}
\mu_{*}=\frac{5}{16\sigma^{2}}\Bigg(\frac{mk_BT}{\pi}\Bigg)^{1/2},\quad \kappa_{*}=\frac{75
 k_B}{16m\sigma^2}\:\Bigg(\frac{mk_BT}{\pi}\Bigg)^{1/2}.
\end{aligned}
\end{equation}

In highly non-equilibrium gas flows, however, the linear constitutive relations are inaccurate. Instead, the exact stress tensor and heat fluxes should encompass not only the linear constitutive relations,  but also high-order terms (HoTs) to account for rarefaction effects. In GSIS, the constitutive relations are constructed as:
\begin{equation}\label{pq_total}
    \begin{aligned}
        &\bm P= (P-\zeta_*{\nabla}\cdot\bm{u})\bm I+ \bm \sigma_{k} + \bm \sigma_{c} = (P-\zeta_*{\nabla}\cdot\bm{u})\bm I+ \bm \sigma_{k}^{\text{NS}} + \bm \sigma_{c}^{\text{NS}} +\text{HoT}_{\bm \sigma},\\
        &\bm q =\bm q_{k}+\bm q_{c}= \bm q_{k}^{\text{NS}}+\bm q_{c}^{\text{NS}}+\text{HoT}_{\bm q},
    \end{aligned}
\end{equation}
where
\begin{equation}\label{DIG-relations}
    \begin{aligned}
        &\text{HoT}_{\bm \sigma} =\bm \sigma_{k}^\text{ESMC} + \bm \sigma_{c}^\text{ESMC} - \bm \sigma_k^{\text{NS}}-\bm \sigma_c^{\text{NS}},\\
        &\text{HoT}_{\bm q}=  \bm q_{k}^\text{ESMC} + \bm q_{c}^\text{ESMC} - \bm q_{k}^{\text{NS}}- \bm q_{c}^{\text{NS}}.
    \end{aligned}
\end{equation}
It should be noted that, the quantities $\bm \sigma_{k}^\text{ESMC}$, $\bm \sigma_{c}^\text{ESMC}$, $\bm q_{k}^\text{ESMC}$, and $\bm q_{c}^\text{ESMC}$ are statistically sampled in the ESMC simulations according to Eq.~\eqref{ESMC_macro}, and $\bm \sigma_{k}^{\text{NS}}$, $\bm \sigma_{c}^{\text{NS}}$, $\bm q_{k}^{\text{NS}}$, $\bm q_{c}^{\text{NS}}$ are calculated according to Eq.~\eqref{NS-relations} with the macroscopic properties $\rho$, $\bm u$, and $T$ sampled from the ESMC as well. However, the NS constitutive relations in Eqs.~\eqref{pq_total} is solved implicitly in the macroscopic synthetic equation. 
Also, it should be emphasized that, the NS constitutive relations in Eqs.~\eqref{pq_total} and~\eqref{DIG-relations} are evaluated at different time steps. The relations in Eq.~\eqref{DIG-relations} are sampled from the stochastic ESMC results at previous time steps, while those in deterministic synthetic equation~\eqref{pq_total} are evaluated at a future time step to guide the evolution of simulation particles. These terms only cancel each other out once the steady state is reached. 

It should be noted that during the ESMC collision process, collision pairs are selected based on a randomly generated unit direction for each particle. For particles located within one diameter from the boundary, the surface facing the boundary is considered protected and is excluded from collision~\cite{frezzotti1999Fourier}. However, the potential part of stress and heat flux contributions from these particles is not precisely sampled by the ESMC method near the boundary. To address the inaccuracy, we estimate these terms using the ratio of the potential to kinetic contributions of the stress and heat flux, denoted as $\mu_c/\mu_k$ and $\kappa_c/\kappa_k$ in Eq.~\eqref{NS-relations}, respectively. All numerical results below show that this is a good practice.

\subsection{DIG Algorithm}

\begin{figure}[t]
	\centering
	\includegraphics[width=1\textwidth, trim=50 70 50 70, clip=true]{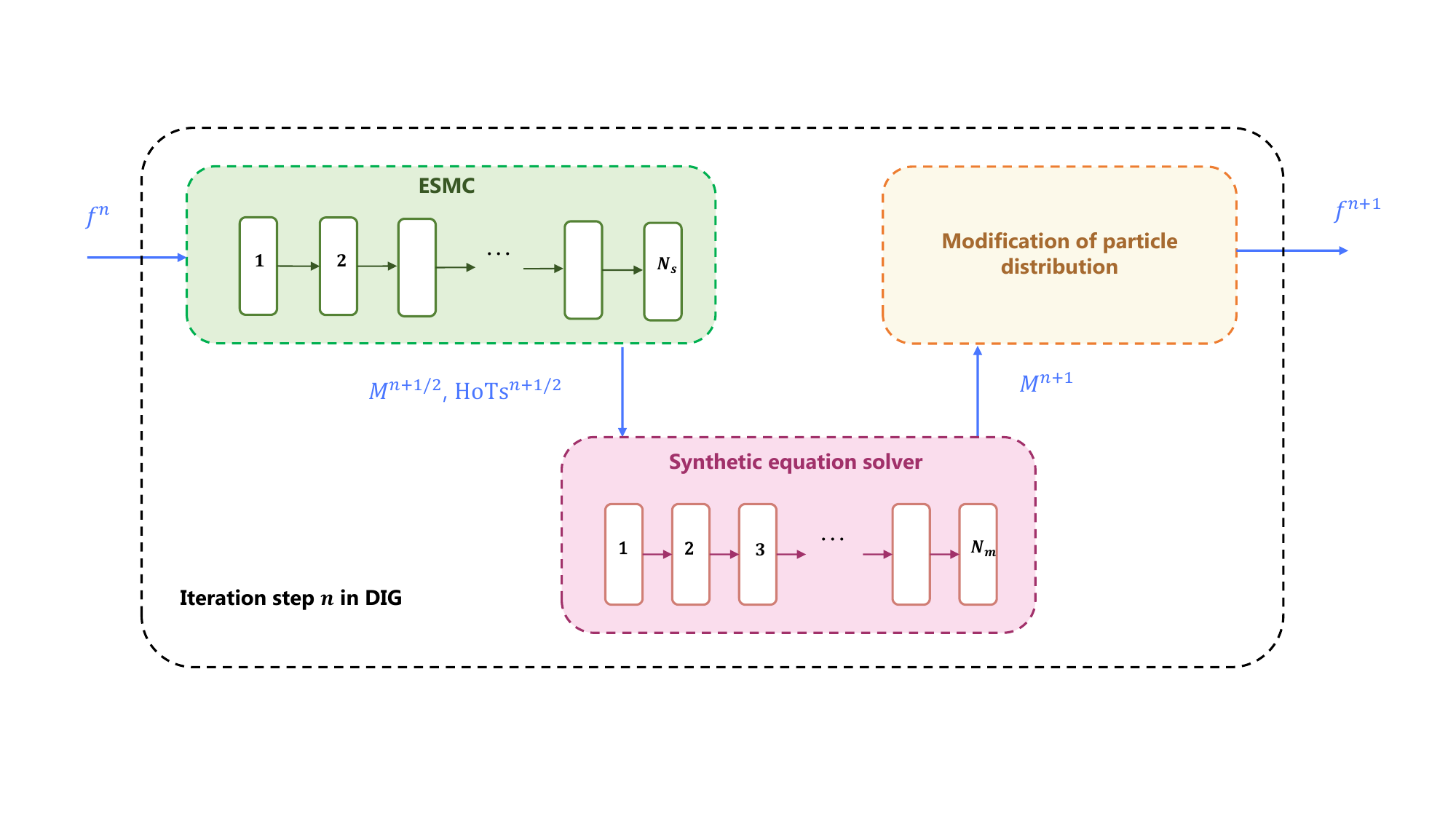}
	\caption{Flowchart of the DIG algorithm for simulating dense gas flows. Each iteration step comprises $N_s$ standard ESMC time steps followed by a solution of the steady-state synthetic equations (with a maximum of $N_m$ inner iterations).} 
	\label{fig:DIG_flowchart}
\end{figure}

The DIG algorithm is proposed to enhance the ESMC simulations.  The main procedure  illustrated in Fig.~\ref{fig:DIG_flowchart} comprises the following steps,
\begin{enumerate}
    \item[1.] Initialization: Solve the macroscopic equation~\eqref{macro} with the NSF constitutive relations~\eqref{NS-relations} only. Initialize the simulation particles based on the equilibrium distribution corresponding to the obtained macroscopic properties.
	
    \item[2.] ESMC solver: In the iteration step $n$, execute the standard ESMC method for $N_s=100$ time steps. Calculate the time-averaged macroscopic quantities $M^{n+1/2}=[\rho^{n+1/2},$ $\bm{u}^{n+1/2}, T^{n+1/2}]$ using Eq.~\eqref{ESMC_macro}, and extract higher-order terms $\text{HoTs}^{n+1/2}$ based on Eq.~\eqref{DIG-relations}. The exponentially weighted moving time average method is adopted to reduce noise for the macroscopic quantities $M^{n+1/2}$, the stress tensor and the heat flux from $\text{HoTs}^{n+1/2}$~\cite{DIG}.
    
    \item[3.] Macroscopic equation solver: Given $M^{n+1/2}$ and $\text{HoTs}^{n+1/2}$, solve the synthetic equations~\eqref{macro} with the full constitutive relations~\eqref{pq_total} for $N_m=500\sim2000$ inner iterations, or until the relative error in macroscopic variables between successive iterations falls below $10^{-5}$. This leads to the updated macroscopic quantities $M^{n+1}$. The boundary condition and the numerical method for solving the synthetic equation are detailed in Refs.~\cite{zeng-2023,liu-2024}.
    
	\item[4.] Particle distribution modification: The DIG is essentially a deterministic–stochastic coupling method, in which the ESMC simulation provides high-order constitutive relations for macroscopic synthetic equations. These synthetic equations, when solved deterministically to steady state, help guide the evolution of the ESMC. Therefore, it is crucial to accurately transfer macroscopic quantities $M^{n+1}$ to the ESMC. Here, the method developed for DSMC is applied~\cite{DIG}.
    
    \item[5.] Repeat Steps 2-4 until the overall solution converges and becomes smooth enough.
\end{enumerate}

\section{Numerical results}\label{sec:5}

Numerical simulations of the one-dimensional normal shock wave, planar Fourier flow and Poiseuille flow, as well as the two-dimensional hypersonic flow past a cylinder and the low-speed porous media flow, are conducted to evaluate the performance of DIG for dense gas flows. In all cases, the gas–surface interactions are assumed to be fully diffuse reflection. The hard-sphere dense gas is argon gas with molecular mass $m=6.63\times10^{-26}$~kg and molecular diameter $\sigma =3.405\times10^{-10}$~m. The macroscopic synthetic equations are computed using the finite volume method, with the conventional fluxes calculated implicitly using the lower-upper symmetric Gauss-Seidel approach. Further implementation details of the numerical scheme are available in Ref.~\cite{gsis_Dense}.  In all our simulations, we set the time steps of the ESMC and the DIG methods to be 0.2 times the required duration of the minimum scale grid distance for particle movement. To ensure the acceleration efficiency of the DIG, the solution of the macroscopic equations is updated every 100 ESMC time steps. Simulations are performed on a parallel computer with the AMD EPYC 7763 processor (2.45~GHz).

\subsection{Normal shock wave}

The simulation of one-dimensional normal shock waves in a dense hard-sphere gas serves as an ideal benchmark to assess the model’s ability to capture both dense-gas behavior and strong non-equilibrium effects, where the influence of boundary condition is absent. The upstream and downstream equilibrium states are specified by the Maxwellian velocity distribution, with the macroscopic quantities satisfy the following Rankine-Hugoniot relations~\cite{frezzotti1993shockwave}:
\begin{equation}\label{RH-shockwave}
    \begin{aligned}
        n_1 u_1 &= n_2 u_2,\\
        n_1 [u_1^2 + RT_1(1+bn_1\chi_1)]&=n_2 [u_2^2 + RT_2(1+bn_2\chi_2)],\\
        u_1^2 + RT_1(5+2bn_1\chi_1) &=  u_2^2 + RT_2(5+2bn_2\chi_2),
    \end{aligned}
\end{equation}
where the variables with subscripts $1$ and $2$ represent the flow field upstream and downstream from the shock, respectively.

\begin{figure}[t!]
\centering
\vspace{-1.5mm}
\subfigure[En = 0 ]{\includegraphics[width=0.45\textwidth,trim=50pt 10pt 50pt 20pt,clip]{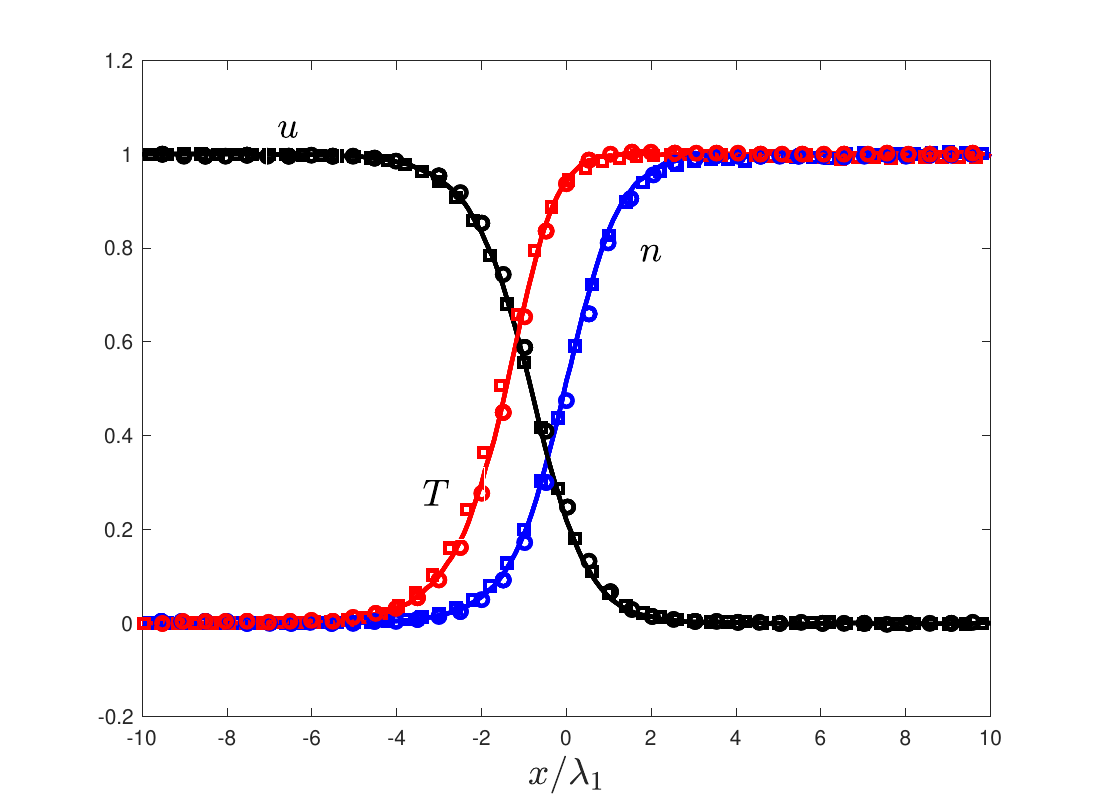}}
 \hspace{0.5cm}
\subfigure[En = 0.4825]{\includegraphics[width=0.45\textwidth,trim=50pt 10pt 50pt 20pt,clip]{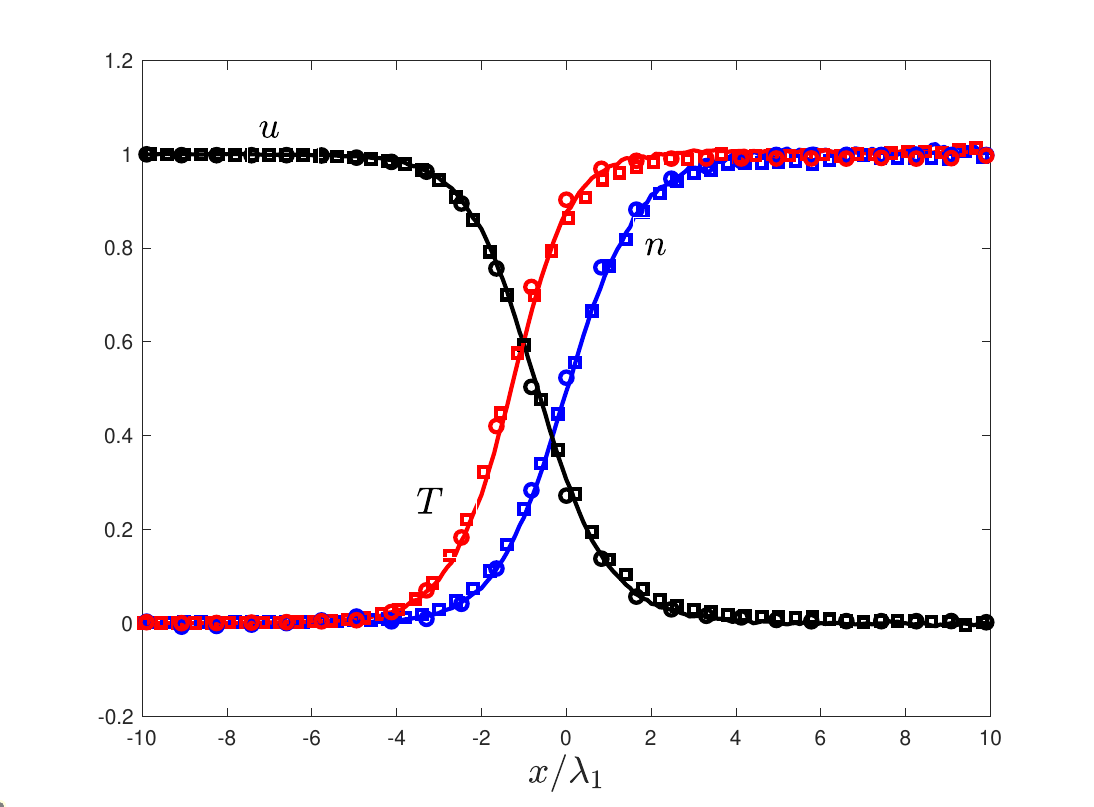}}
\caption{The normalized $n$, $u$, and $T$ obtained from the ESMC  (line) and DIG (squares) of the Enskog equation, as well as the reference solution (circles) from the MD simulation~\cite{frezzotti1998molecular}, when Ma = 4.
Reference values are assigned using the upstream density $n_1$ and temperature $T_1$, with flow velocity normalized by $v_0=\sqrt{2R T_1}$. The $x$-axis is shifted so that $n(x=0)=(n_1+n_2)/2$. The macroscopic quantities $W=\{n, u, T\}$ are further normalized as $(W-\text{min}(W_1, W_2))/|W_1-W_2|$, respectively.
} %% tecplot 重新画

\label{fig: ESMC_DIG_Ref}
\end{figure}

The normal shock wave in dense gases is characterized by two key parameters~\cite{frezzotti1993shockwave}: the Mach number $\text{Ma} = u_1/\sqrt{\gamma RT_1}$ and the upstream Enskog number  $\text{En}= \sigma / \lambda_1$, with $\gamma = 5/3$ being the specific heat ratio for monatomic gases and $\lambda_1 = 1/[\sqrt{2}\pi \sigma^2n_1 \chi(n_1)]$ being the upstream mean free path. In our simulations, the computational domain is set to be $[-50\lambda_1,50\lambda_1]$, which is uniformly meshed by 1000 points. 

Figure~\ref{fig: ESMC_DIG_Ref} compares the normalized density, velocity, and temperature profiles obtained from the ESMC, DIG, and the reference solutions from the molecular dynamics simulation~\cite{frezzotti1998molecular}. Excellent agreement is observed across the shock structure at Ma = 4 for both dilute (En = 0) and dense (En = 0.4825) gas conditions, confirming the validity of the ESMC and DIG in solving the Enskog equation. In the dilute limit (En = 0), the Enskog equation reduces to the Boltzmann equation, and the ESMC simplifies to the DSMC method. Under these conditions, both ESMC and DIG remain consistent with the reference. Moreover, as the gas density increases to En=0.4825, the solutions continue to agree well with molecular dynamics results, demonstrating that the ESMC and DIG are robust not only in dilute gases but also in dense gases. Therefore, in what follows, the solutions of the Enskog equation obtained via the ESMC will serve as the benchmark for validating the DIG.

% Figure \ref{fig: ESMC_DIG_shockwave_iteration_speed} illustrates the evolution of the velocity profiles of the shock wave when En = 0.2. The ESMC  requires about 3000 steps to reach a comparable level of convergence. In contrast, the DIG achieves excellent agreement with the reference 1000 iterative steps. 

% \begin{figure}
%     \centering
%     \includegraphics[width=0.49\textwidth]{figures/Kn0.010_Ma4.0_Eta0.022_ESMC_shockwave_step.eps}
%     \includegraphics[width=0.49\textwidth]{figures/Kn0.010_Ma4.0_Eta0.022_DIG_shockwave_step.eps}
%     % \includegraphics[width=0.49\textwidth]{figures/Kn0.010_Ma4.0_Eta0.050_ESMC_shockwave_step.eps}
%     % \includegraphics[width=0.49\textwidth]{figures/Kn0.010_Ma4.0_Eta0.050_DIG_shockwave_step.eps}
%     \caption{The comparison of the normalized velocity profiles at different time steps between ESMC  and DIG, when En = 0.2 and Ma = 4.}
%     \label{fig: ESMC_DIG_shockwave_iteration_speed}
% \end{figure}

\subsection{Planar Fourier flow}

%Upon collision, a molecule thermalizes with the wall: its incident velocity is discarded and a new velocity is drawn from the Maxwell–Boltzmann distribution at the wall’s temperature. 
%Consequently, the gas is driven out of equilibrium, sustaining a temperature gradient and a steady heat flux. Because the wall screens collisions through the Enskog operator, a dense adsorption layer forms at each surface—an effect wholly absent in the dilute-gas Boltzmann description.

\begin{figure}[t!]
    \centering
     \includegraphics[width=0.43\textwidth, trim=80pt 40pt 120pt 50pt,clip]{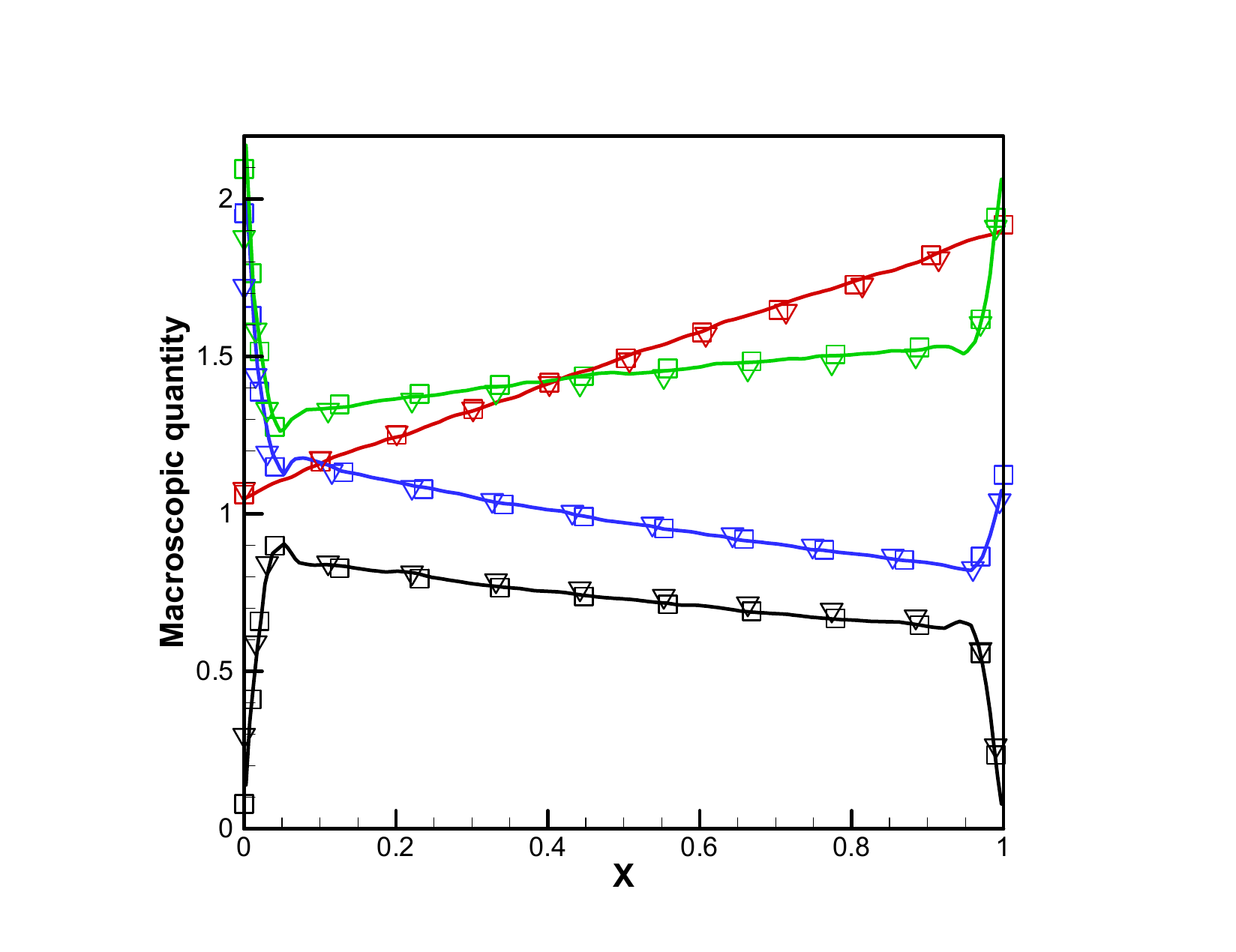}
    \hspace{0.5cm}
    \includegraphics[width=0.43\textwidth, trim=80pt 40pt 120pt 50pt,clip]{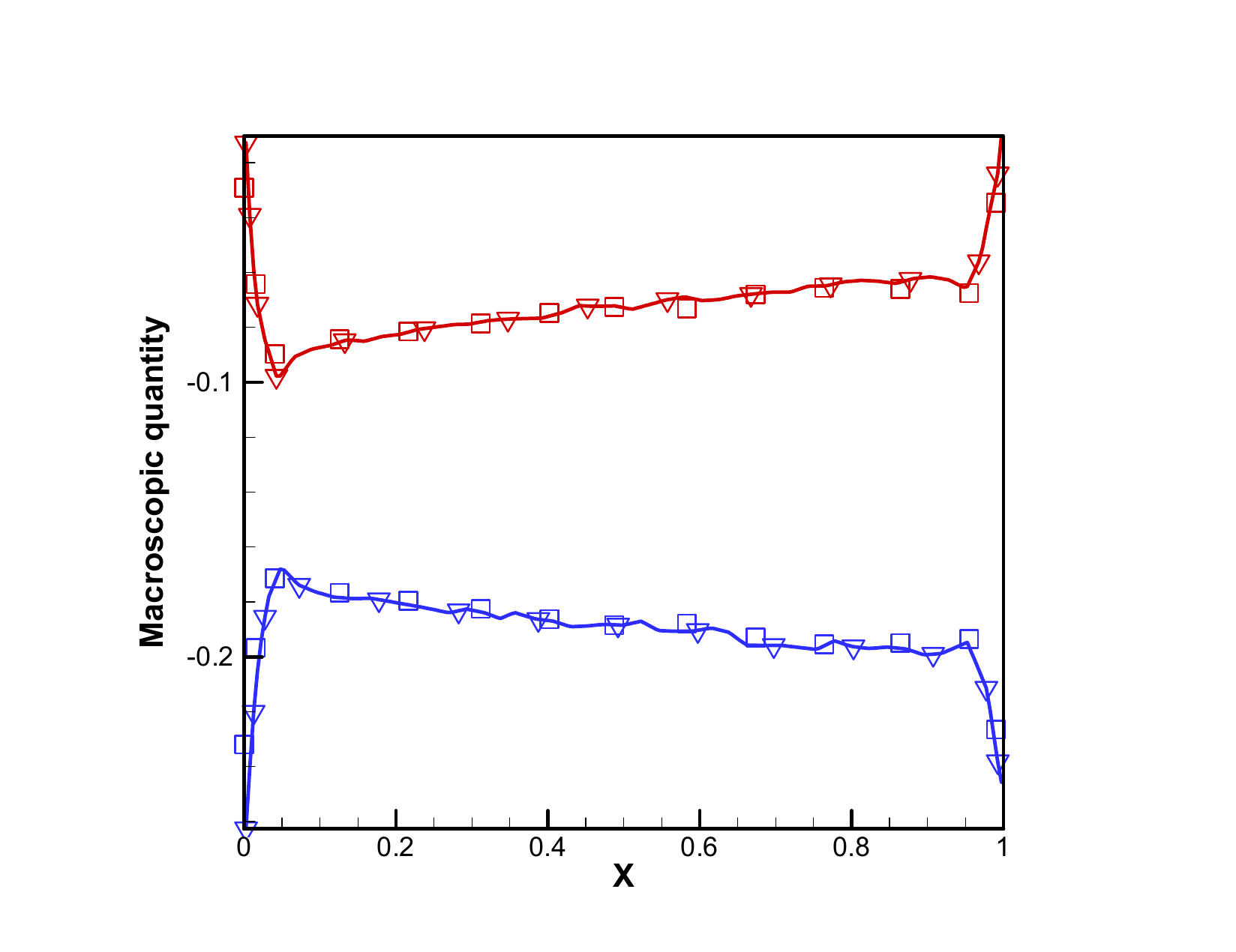}
    \\
    \includegraphics[width=0.43\textwidth, trim=80pt 40pt 120pt 50pt,clip]{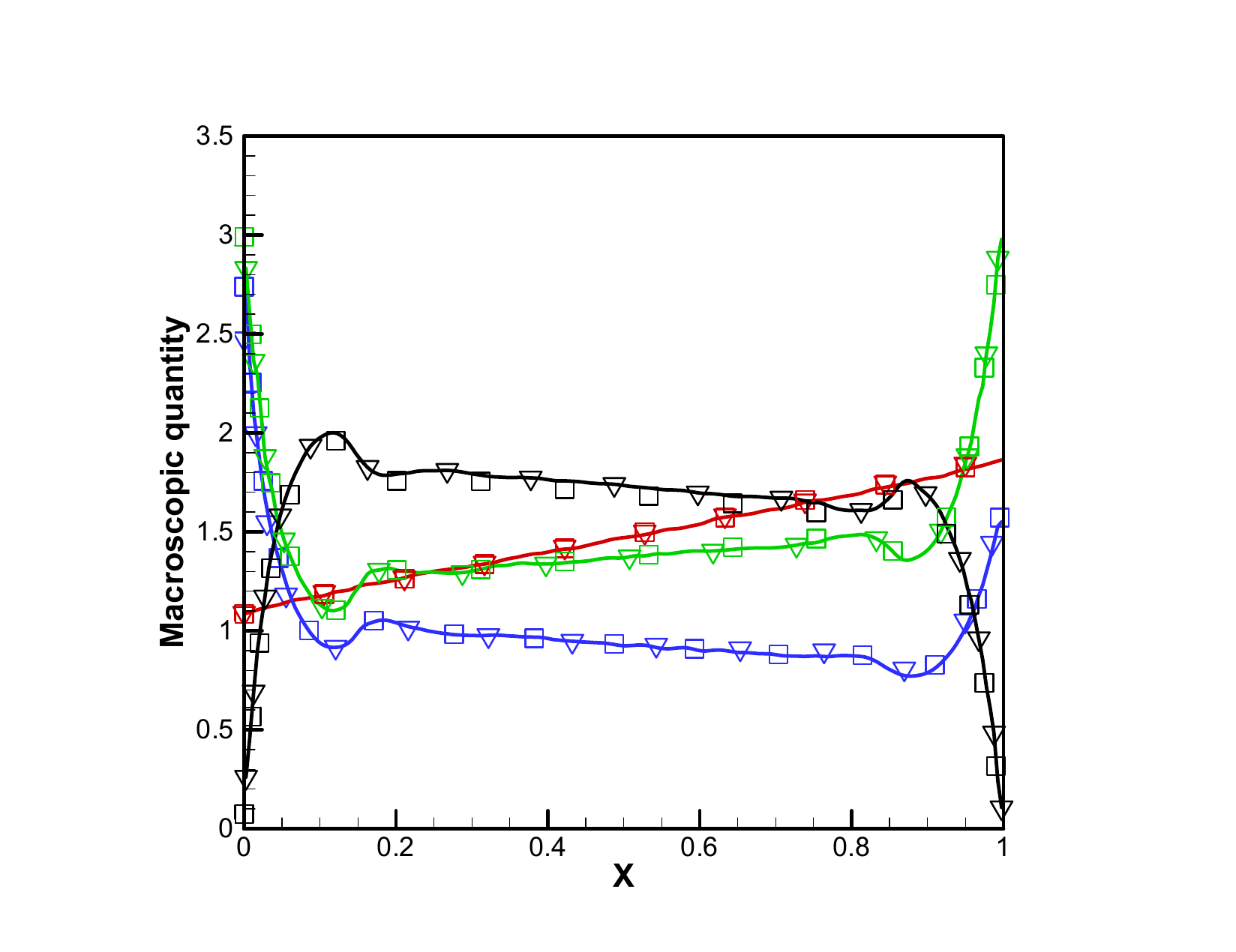}
    \hspace{0.5cm}
    \includegraphics[width=0.43\textwidth, trim=80pt 40pt 120pt 50pt,clip]{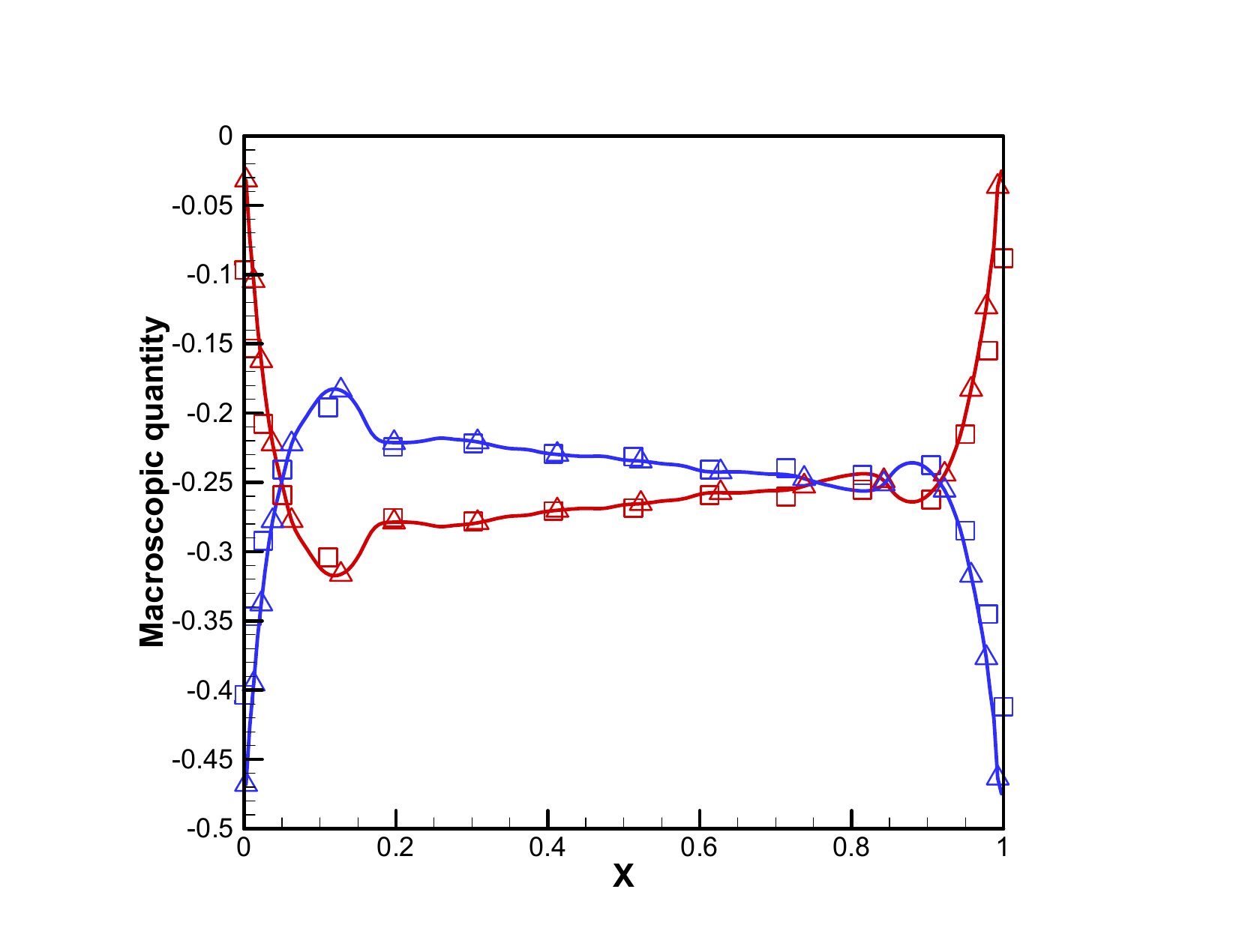}
    \caption{Comparison of ESMC (squares), DIG (triangles), and the reference (lines) solutions of the Enskog equation for the Fourier flows of hard-sphere molecules with $T_2=2T_1$. The Knudsen number is Kn = 0.05, and the Enskog numbers in the first and second rows are En = 1.106 and 2.983, respectively. The macroscopic quantities in the left column are density (blue), temperature (red), as well as the kinetic (green) and potential (black) parts of pressure, while the quantities in the right column are the kinetic (blue) and potential (red) parts of heat flux. }
    \label{fig: Fourier_ESMC_DIG_comparsion}
\end{figure}

The one-dimensional Fourier flow provides a good test case for assessing the accuracy of DIG in handling wall-confined flows. In this case, hard-sphere molecules of diameter $\sigma$ are confined between two plates at $x=-\frac{\sigma}{2}$ and $x=L+\frac{\sigma}{2}$. The left plate is held at a temperature $T_1$, the right plate at $T_2>T_1$. The one-dimensional spatial domain is discretized into 400 uniform cells in ESMC, while the DIG achieves comparable accuracy to ESMC with 100 cells.
%demonstrating certain improved computational efficiency and asymptotic-preserving property.  

Figure \ref{fig: Fourier_ESMC_DIG_comparsion} presents the density $n$, temperature $T$, and the pressure $P_{xx}$ and heat flux $q_x$ of the Fourier flow predicted by the ESMC, DIG, and the reference solutions of the Enskog equation~\cite{frezzotti1999Fourier}; these reference solutions are also validated by the molecular dynamics simulation. 
The $x$-axis is normalized by $L$, the density, velocity, temperature, pressure, and heat flux are normalized by $n_0$, $T_1$, $\sqrt{RT_1}$, $n_0k_BT_1$, and $n_0k_BT_1\sqrt{RT_1}$, respectively. The Knudsen number is defined in Eq.~\eqref{mfp} based on the average number density $n_0$.
The ESMC results exhibit excellent agreement with the reference solution, thereby validating again the accuracy of ESMC. In contrast to dilute gas, the density profile of dense gas no longer follows a monotonically decreasing function along the spatial coordinate. Instead, it exhibits oscillatory behavior near the cold wall, with fluctuations occurring on the scale of the molecular diameter $\sigma$. As the gas density increases, the amplitude of these peaks becomes more pronounced. The elevated density profiles observed near the boundary can be attributed to a volume exclusion effect: when the distance between a molecule and the wall is less than the molecular diameter, a portion of the molecular surface is excluded from collisions due to the lack of sufficient space for collision partners. This effect effectively pushes molecules toward the wall, leading to local accumulation and thus an increase in density.

Due to the nonlocal collision, the potential parts of stress and heat flux exist. When the steady state is reached, the total pressure $P_{xx}$ and heat flux $q_x$ are constant over the computational domain, which can be simplified from the macroscopic synthetic equations, and seen in Fig.~\ref{fig: Fourier_ESMC_DIG_comparsion}.

% \begin{figure}[t!]  
%     \centering
%     \includegraphics[width=0.4\textwidth,trim=80pt 30pt 120pt 50pt,clip]{figures/Poiseuille_ESMC_DIG_density_Kn0.1.pdf} 
%     \includegraphics[width=0.4\textwidth,trim=80pt 30pt 120pt 50pt,clip]{figures/Poiseuille_ESMC_DIG_density_Kn0.01.pdf}
%     \\
%     \includegraphics[width=0.4\textwidth,trim=80pt 30pt 120pt 50pt,clip]{figures/Poiseuille_ESMC_DIG_velocity_Kn0.1.pdf}
%     \includegraphics[width=0.4\textwidth,trim=80pt 30pt 120pt 50pt,clip]{figures/Poiseuille_ESMC_DIG_velocity_Kn0.01.pdf}
%     \\
%     \includegraphics[width=0.4\textwidth,trim=80pt 30pt 120pt 50pt,clip]{figures/Poiseuille_ESMC_DIG_Temperature_Kn0.1.pdf}
%     \includegraphics[width=0.4\textwidth,trim=80pt 30pt 120pt 50pt,clip]{figures/Poiseuille_ESMC_DIG_Temperature_Kn0.01.pdf}
%     \caption{The density, velocity, and temperature profiles in the force-driven Poiseuille flow. The Knudsen numbers in the left and right columns are 0.1 and 0.01, respectively.  H500, H100, and H50 mean that the computational domain is discretized uniformly by 500, 100, and 50 cells, respectively. 
%     }
%     \label{Poiseuille_figure}
% \end{figure}

\subsection{Planar force-driven Poiseuille flow}

The force-driven Poiseuille flow is an excellent test case for both the fast-converging and asymptotic-preserving properties of DIG, since at small Kn the conventional scheme converges very slowly and requires a large number of spatial grids~\cite{wang2018comparative}. The one-dimensional flow is confined between two infinite plates maintained at temperature $T_0$, positioned at $x=-\frac{\sigma}{2}$ and $x=L+\frac{\sigma}{2}$. The initial average number density $n_0$ is tuned so that  Kn = 0.01 and 0.1 are considered, coupled with the Enskog numbers En = 0.01 and 0.5. A uniform external force is applied in the $y$ direction, with the dimensionless acceleration 
$\text{Fr} = \frac{m a L}{2k_BT_0}$. The values of Fr are set to be 0.05 and 0.5 for Kn = 0.01 and 0.1, to keep the peak velocity around one Mach.

\begin{figure}[t!]  
    \centering
    \includegraphics[width=0.32\textwidth,trim=80pt 30pt 120pt 50pt,clip]{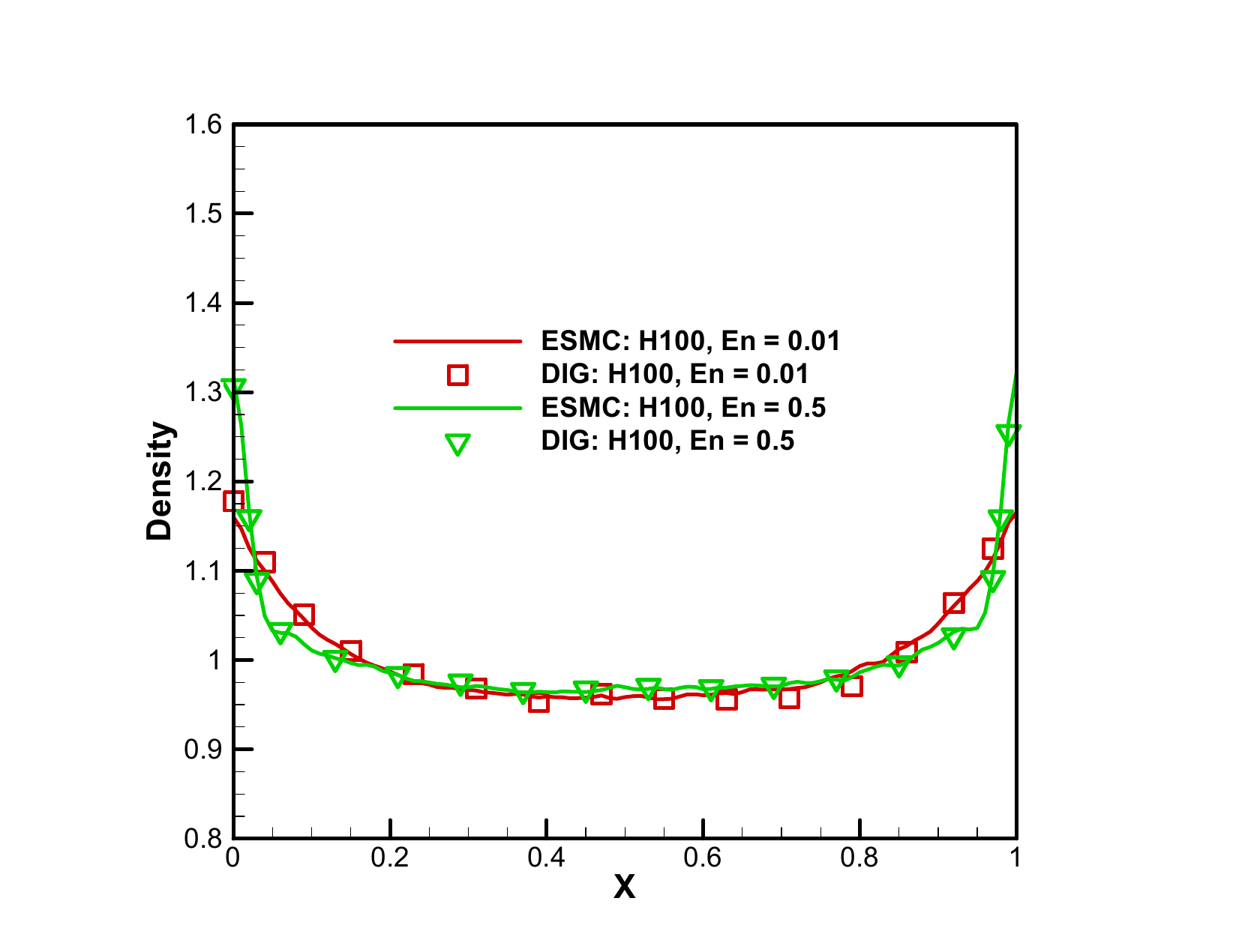} 
\includegraphics[width=0.32\textwidth,trim=80pt 30pt 120pt 50pt,clip]{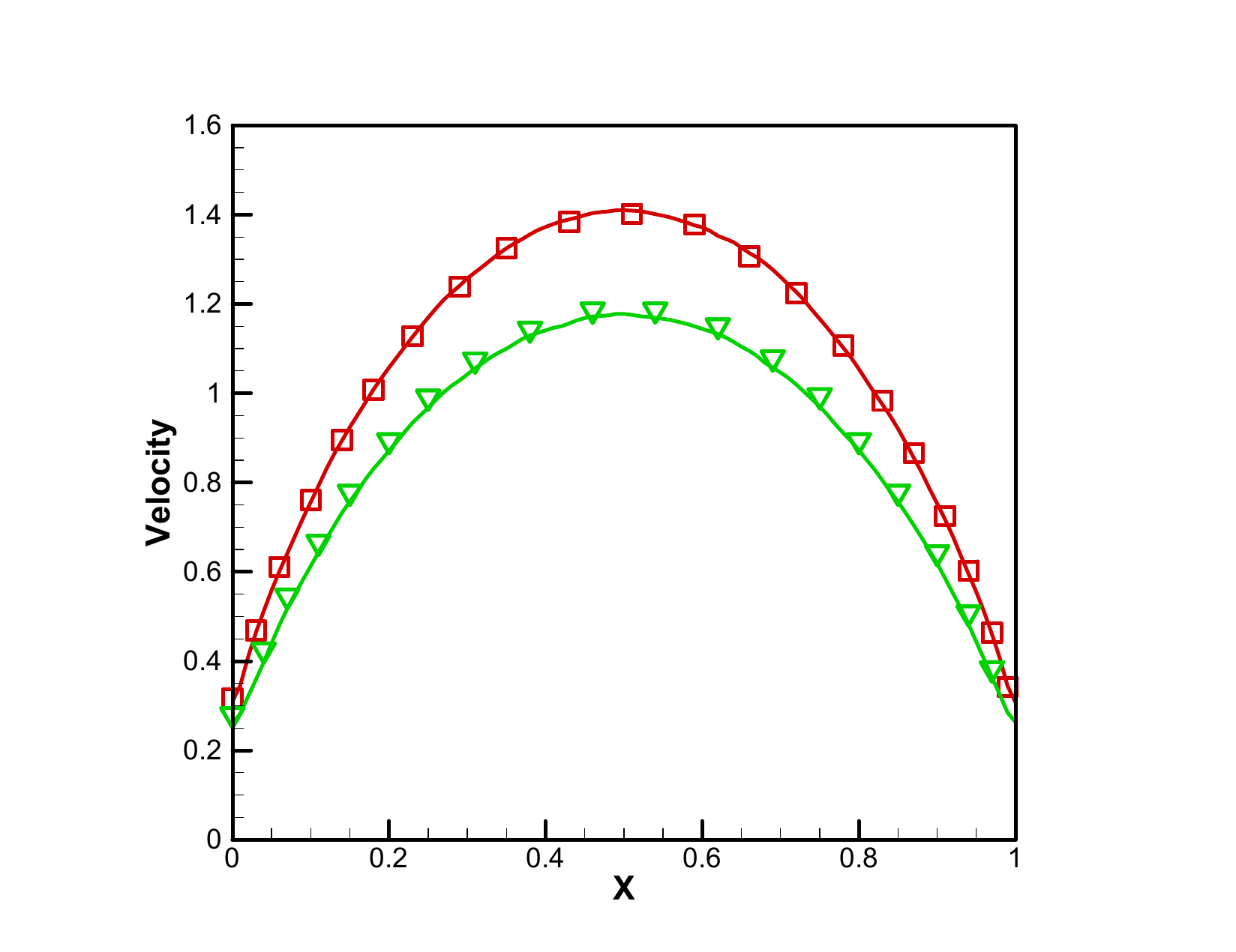}
    \includegraphics[width=0.32\textwidth,trim=80pt 30pt 120pt 50pt,clip]{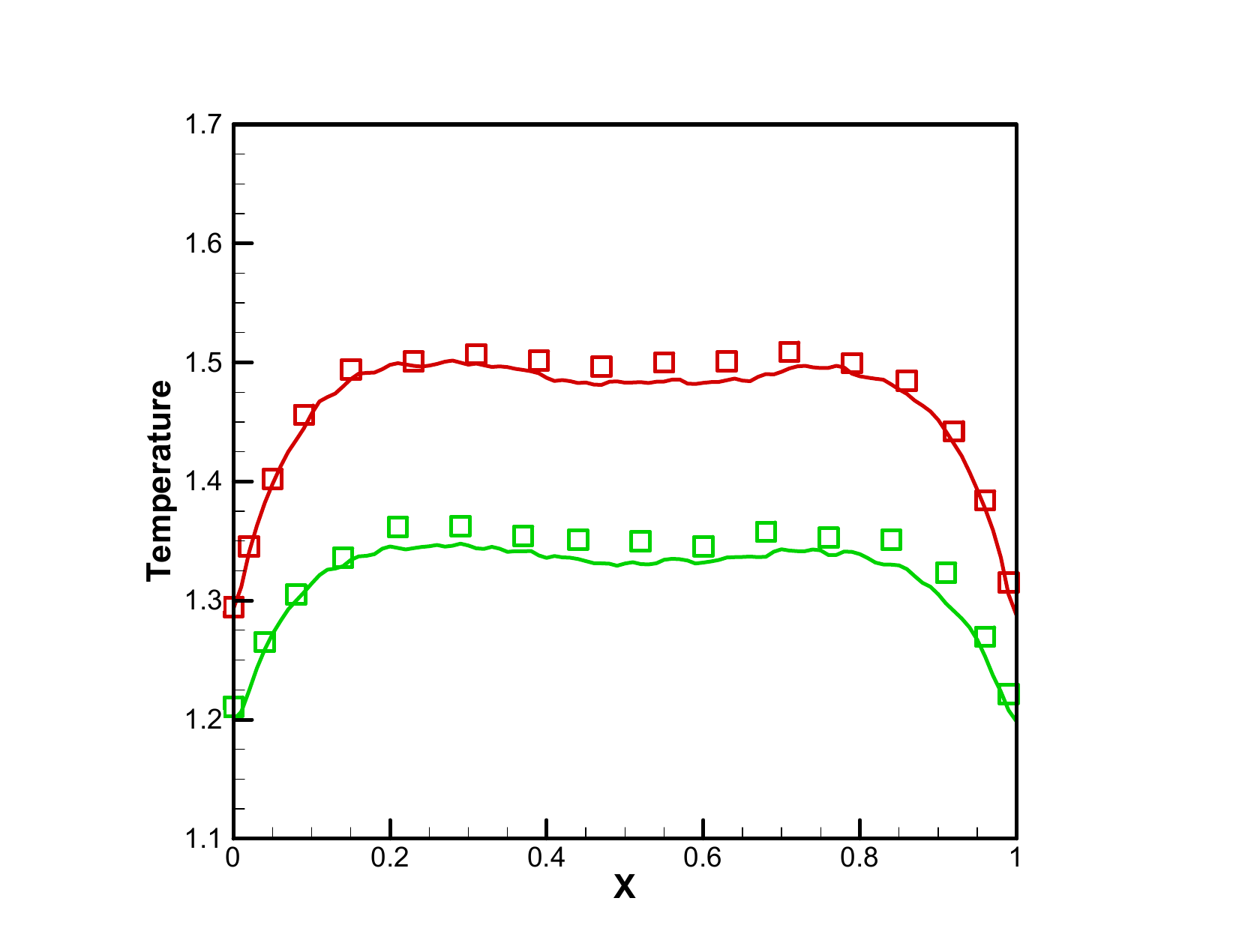}
    \\
    \includegraphics[width=0.32\textwidth,trim=80pt 30pt 120pt 50pt,clip]{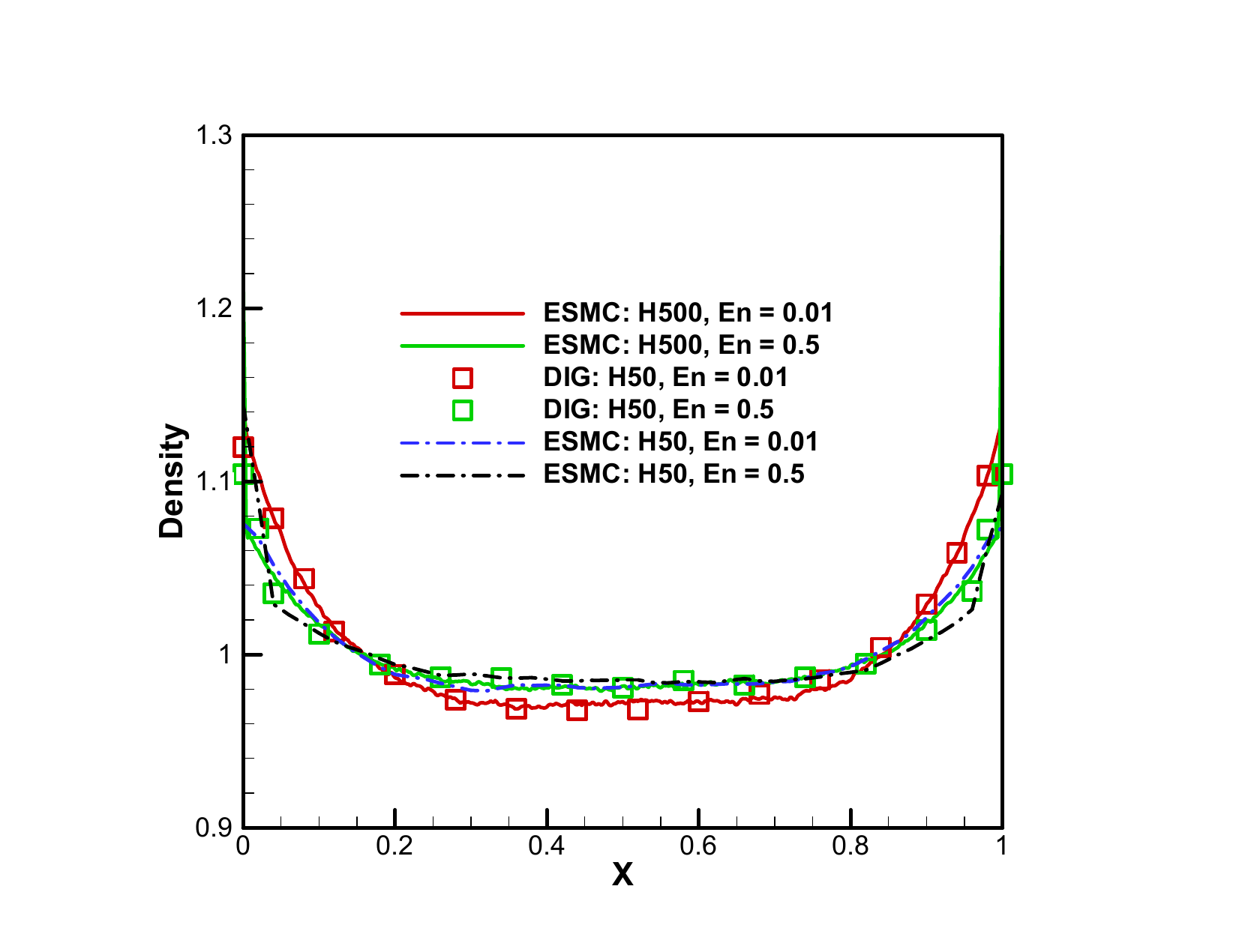}
    \includegraphics[width=0.32\textwidth,trim=80pt 30pt 120pt 50pt,clip]{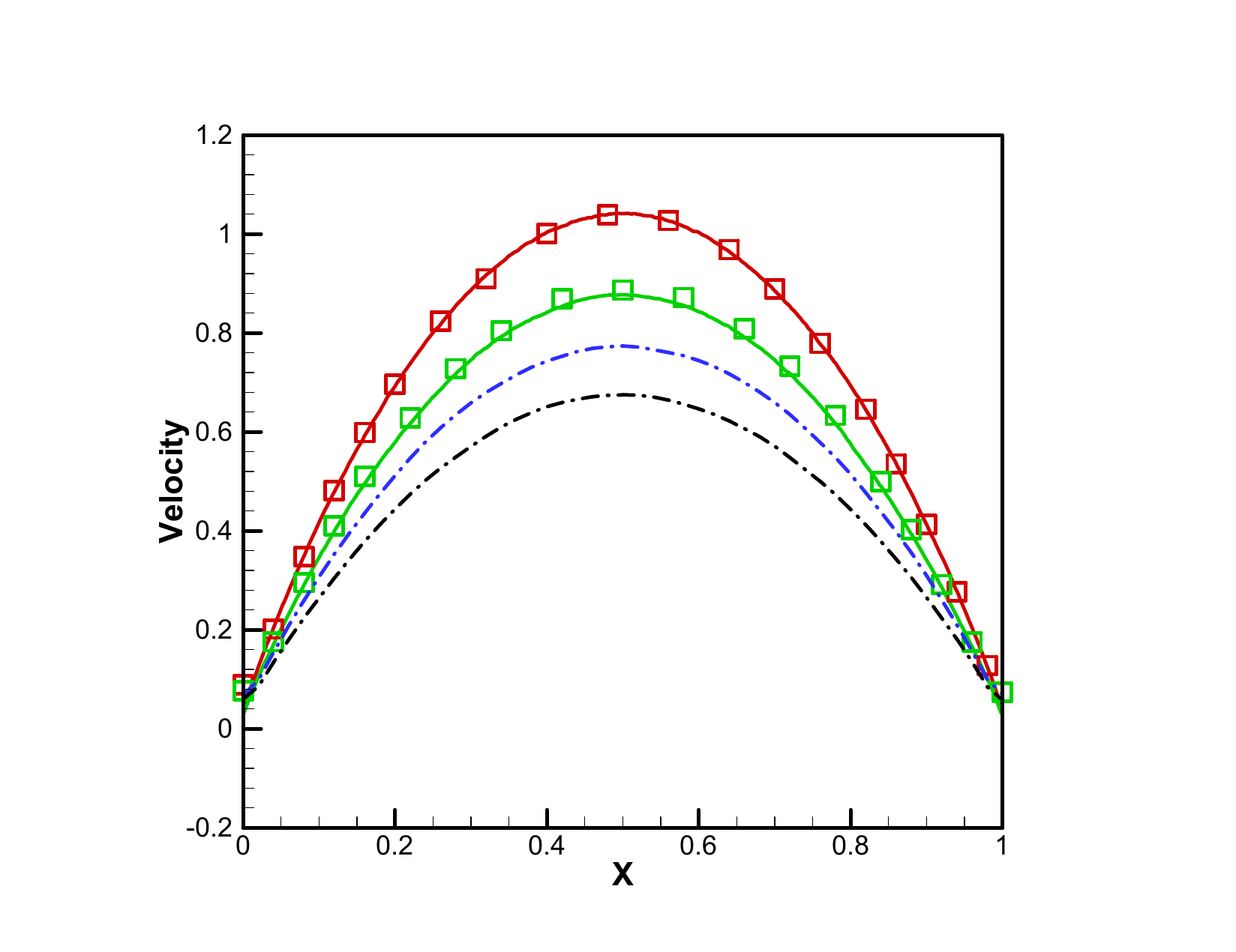}
    \includegraphics[width=0.32\textwidth,trim=80pt 30pt 120pt 50pt,clip]{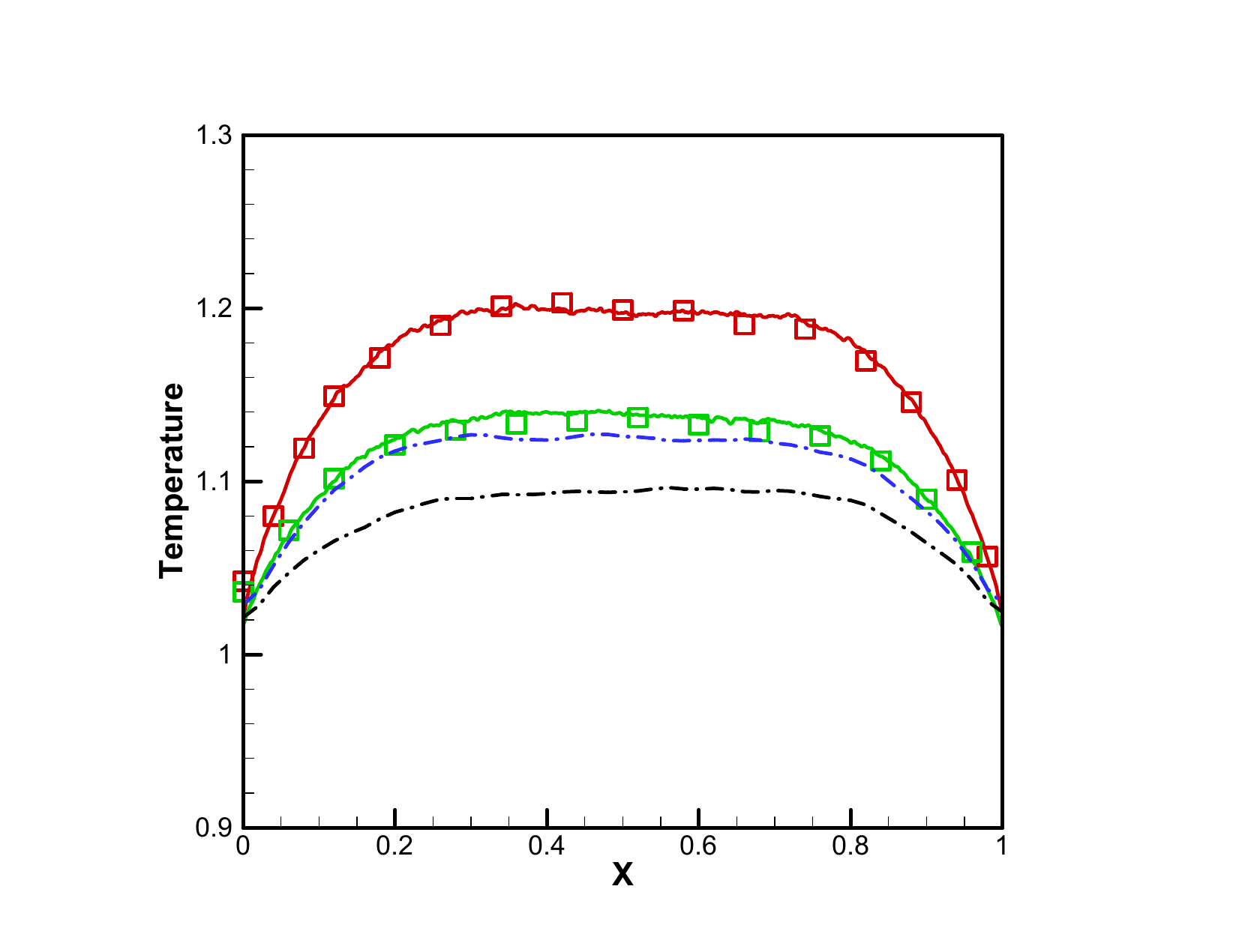}
    \caption{The density, velocity, and temperature profiles in the force-driven Poiseuille flow. The Knudsen numbers in the first and second rows are 0.1 and 0.01, respectively.  H500, H100, and H50 mean that the computational domain is discretized uniformly by 500, 100, and 50 cells, respectively. 
    }
    \label{Poiseuille_figure}
\end{figure}

The density, velocity, and temperature profiles are shown in Fig.~\ref{Poiseuille_figure}. First, due to the volume exclusion effect, the density in the vicinity of solid walls increases with the Enskog number. Second, when the Knudsen number is fixed, the velocity profile becomes flatter as En increases. This is because the viscosity~\eqref{mu_kap} is not only proportional to $1/\chi$,  but also to  $(1+\frac{2}{5}bn\chi)$; the latter increases with En. Larger viscosity means larger physical dissipation, and hence the peak velocity decreases when En increases. On the other hand, when En is determined, the velocity slip at the boundary becomes more pronounced with the increasing Kn, i.e., the velocity slip roughly increases from 0.05 to 0.3 when Kn increases from 0.01 to 0.1.
Third, in continuum thermodynamics, viscous dissipation would make the gas temperature highest at the channel centerline, leading to a unimodal parabolic temperature profile. However, in force-driven Poiseuille flow, kinetic theory predicts a bi-modal temperature distribution, i.e., the temperature is lower at the centerline, and two off-center peaks appear near the walls. This non-intuitive behavior is a strong signature of rarefaction effects, which violate the linear constitutive relations, even in dense gases. It is noted that such behaviors also exist in dilute rarefied gas flows~\cite{zhengcomparison_poise} and dense gas flows predicted by the event-driven molecular dynamic simulations~\cite{shen2010HD}. When the Knudsen number is decreased from 0.1 to 0.01, the bi-modal profile becomes less apparent due to the reduction of rarefaction effects.

For the traditional ESMC method, the spatial cell size must be approximately one-third of the mean free path, and the time step should be on the order of one-third of the mean collision time. Consequently, an enormous amount of iterations is required to reach a steady state, followed by extensive sampling to obtain the stable velocity distribution. For example, Fig.~\ref{Poiseuille_figure} shows that when Kn=0.01, using only 50 grid cells results in an underestimation of the flow temperature and velocity, due to the large numerical dissipation. In sharp contrast, owing to its incorporation of macroscopic synthetic equation, the DIG accurately captures the density, velocity, and temperature distributions, demonstrating a good asymptotic-preserving property.
Furthermore, since in the DIG the macroscopic synthetic equations are solved towards the steady state, it helps the simulation particles quickly evolve to the steady state. For instance, when Kn = 0.01 and En = 0.01, the ESMC requires approximately 400,000 iterations and 104 minutes of OpenMP parallelization with 40 cores to achieve steady state, while DIG converged within only 4,000 steps and 1.3 minutes with the same cores, representing an approximate 100-fold reduction in iteration count and computational time. This provides strong evidence of the fast-converging property of DIG. The gain in computational efficiency of DIG would be even larger when the Knudsen number further decreases.

\begin{figure}[t!]
\centering
 \includegraphics[width=0.44\textwidth, trim=80pt 20pt 100pt 80pt,clip ]{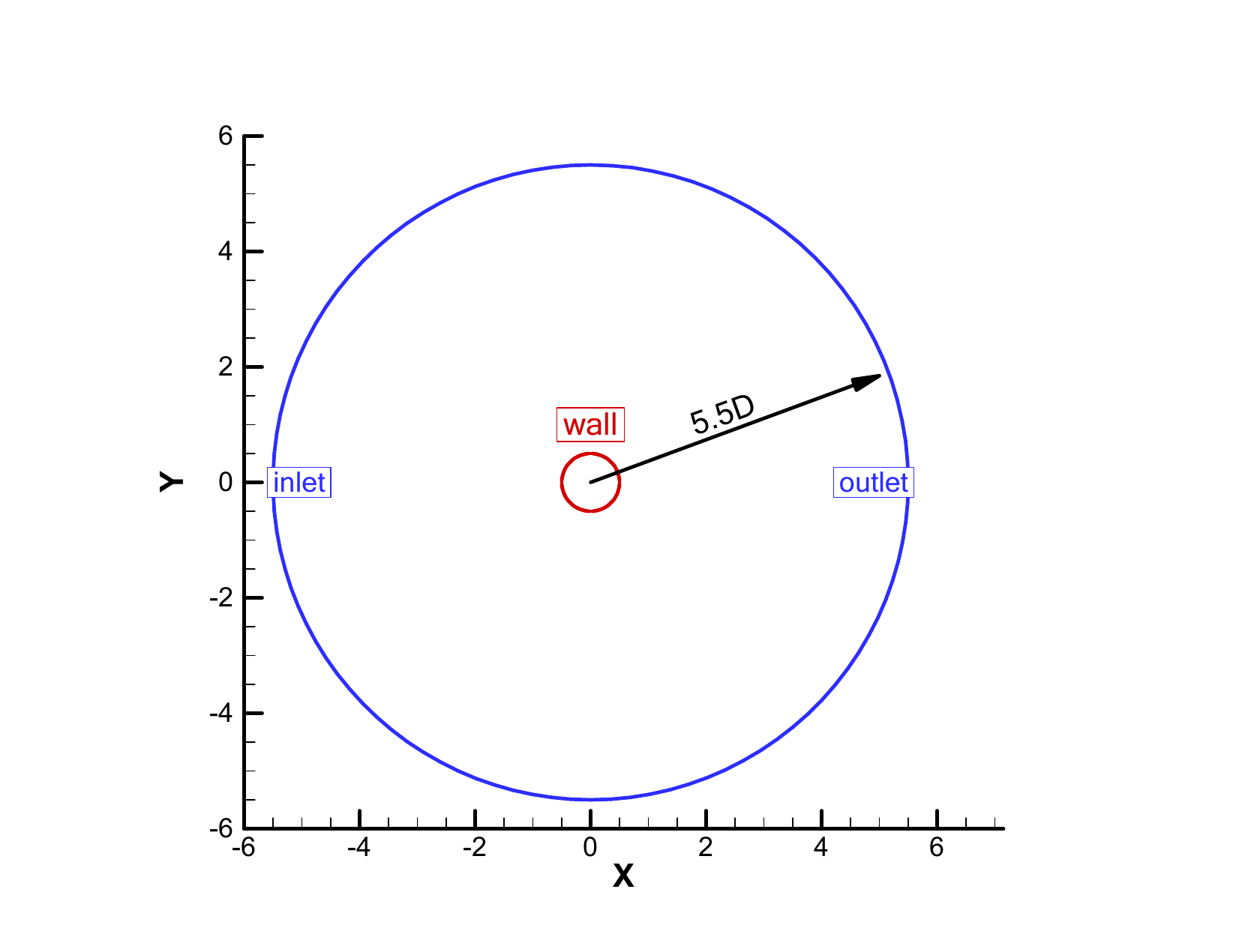}
 \includegraphics[width=0.44\textwidth,trim=80pt 30pt 100pt 85pt, clip ]{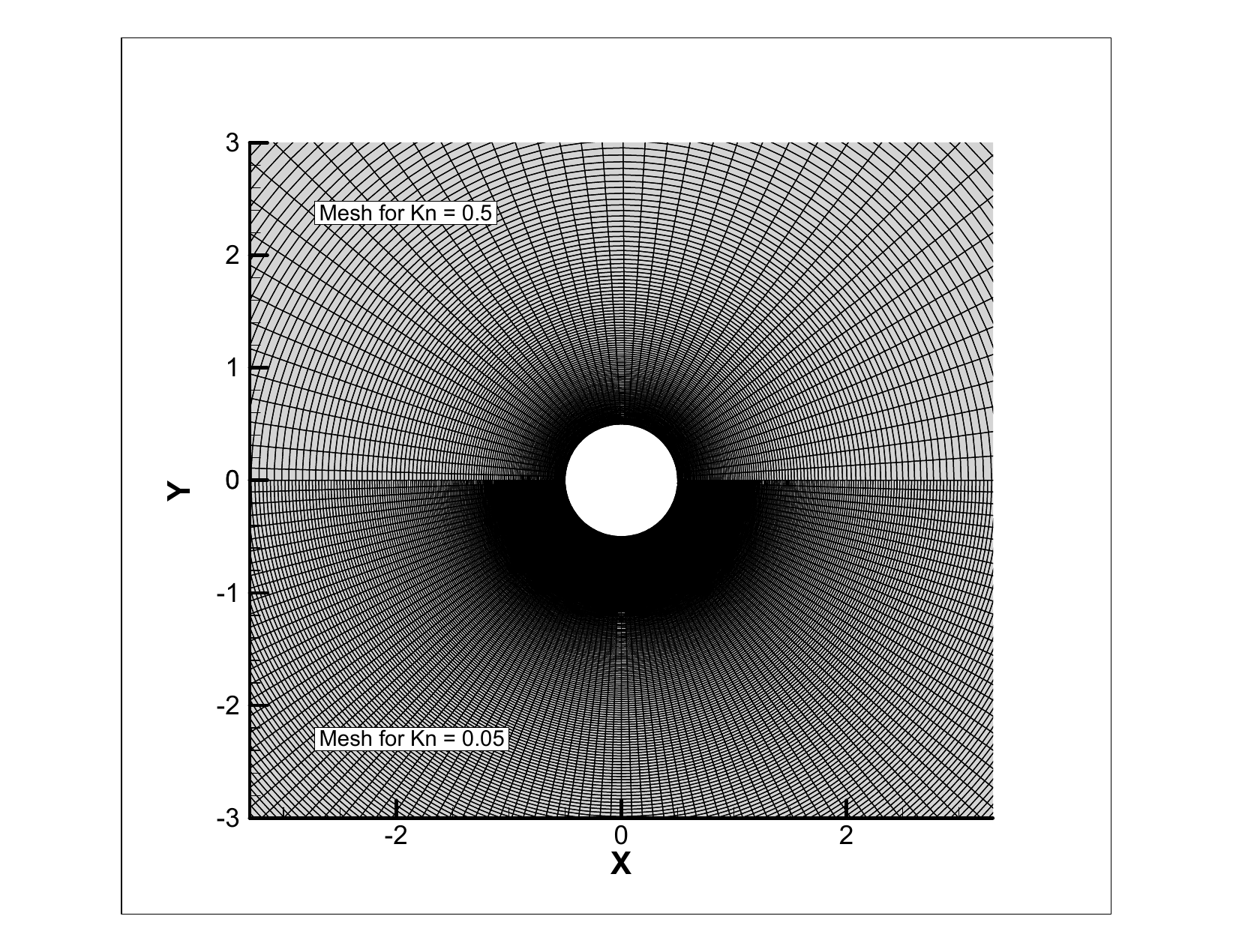}
        \\
\includegraphics[width=0.32\textwidth,trim=100pt 40pt 100pt 40pt,clip]{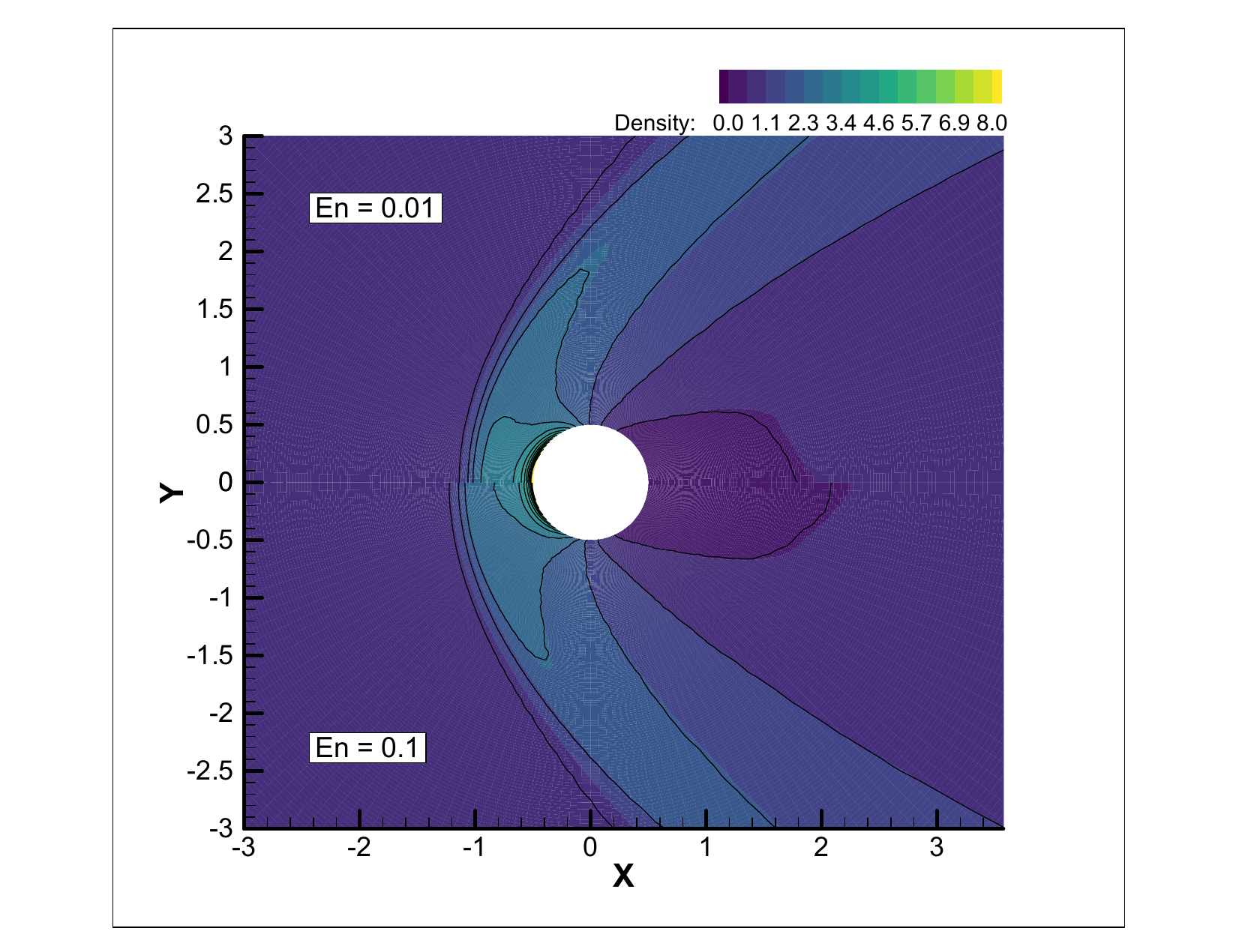}
\hfill
\includegraphics[width=0.32\textwidth,trim=100pt 40pt 100pt 40pt,clip]{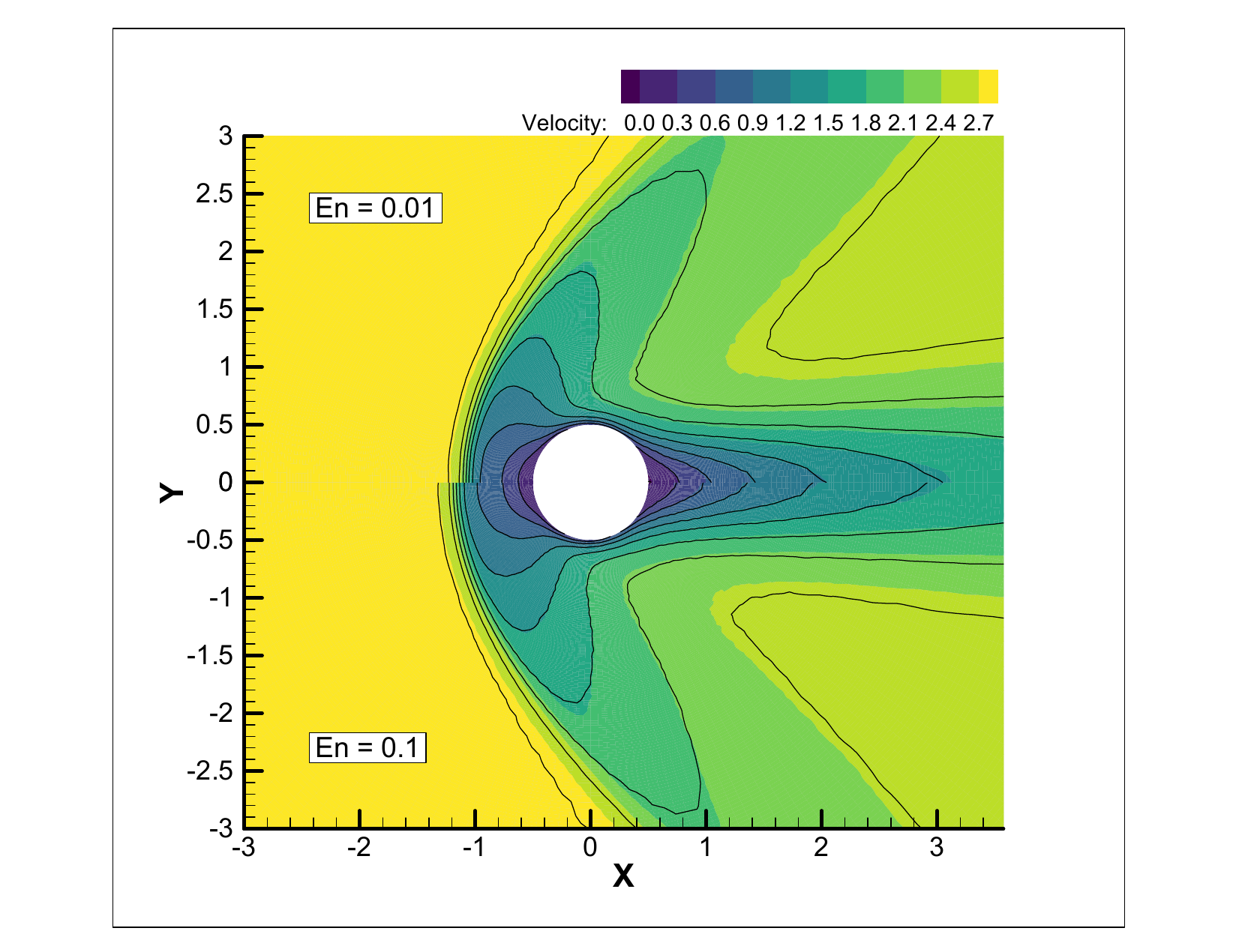}
\hfill
\includegraphics[width=0.32\textwidth,trim=100pt 40pt 100pt 40pt,clip]{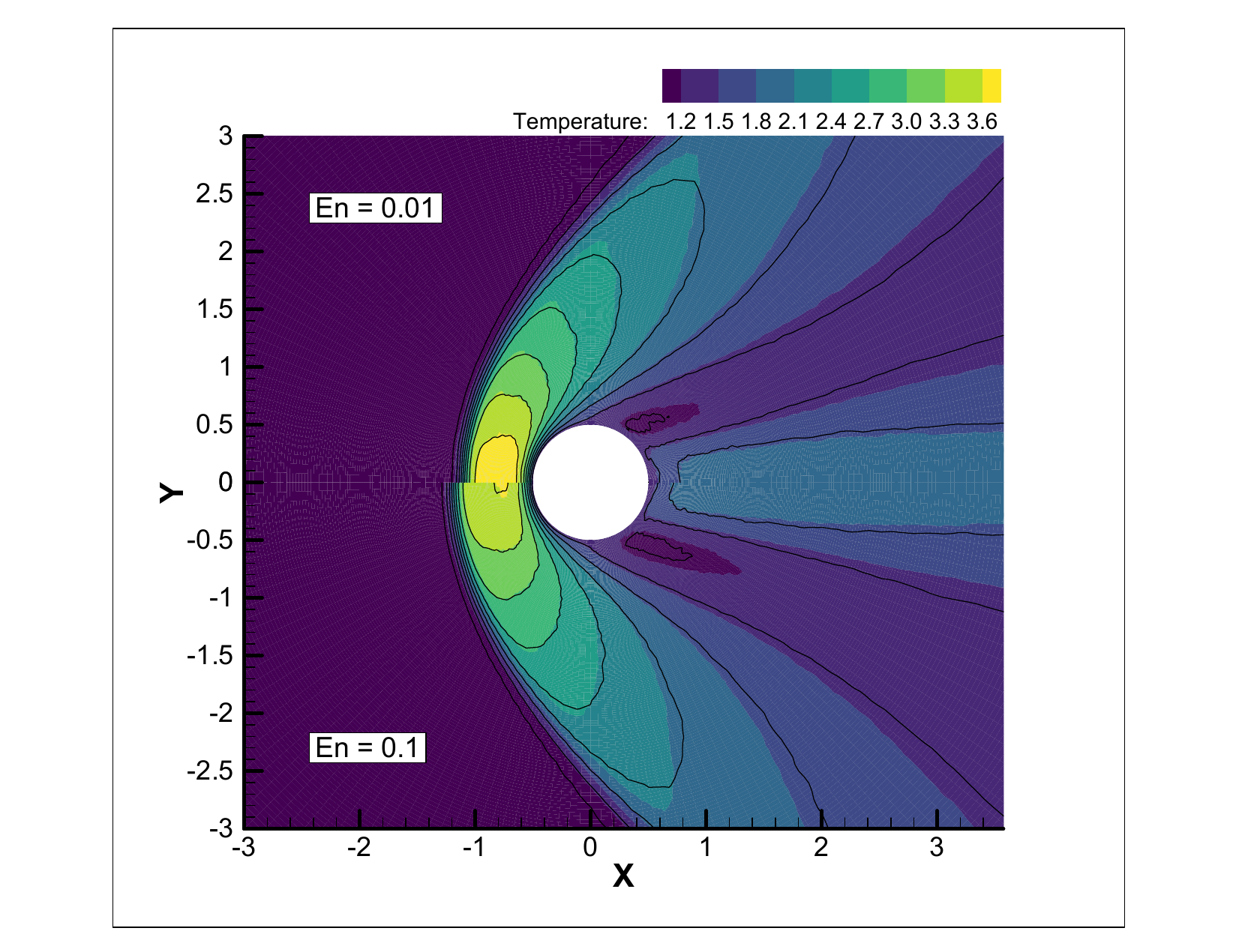}
\\
\includegraphics[width=0.32\textwidth,trim=100pt 40pt 100pt 40pt,clip]{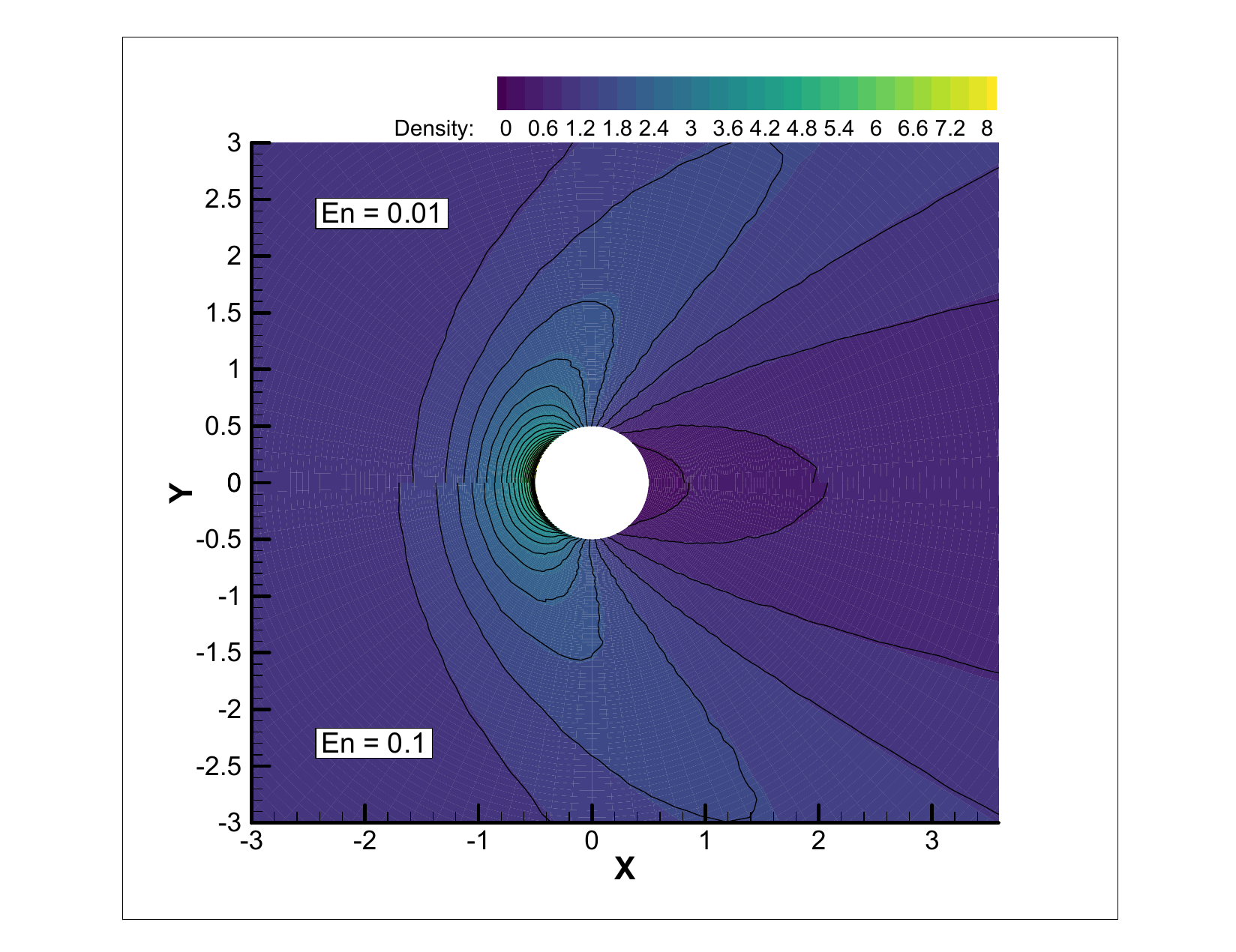}
\hfill
\includegraphics[width=0.32\textwidth,trim=100pt 40pt 100pt 40pt,clip]{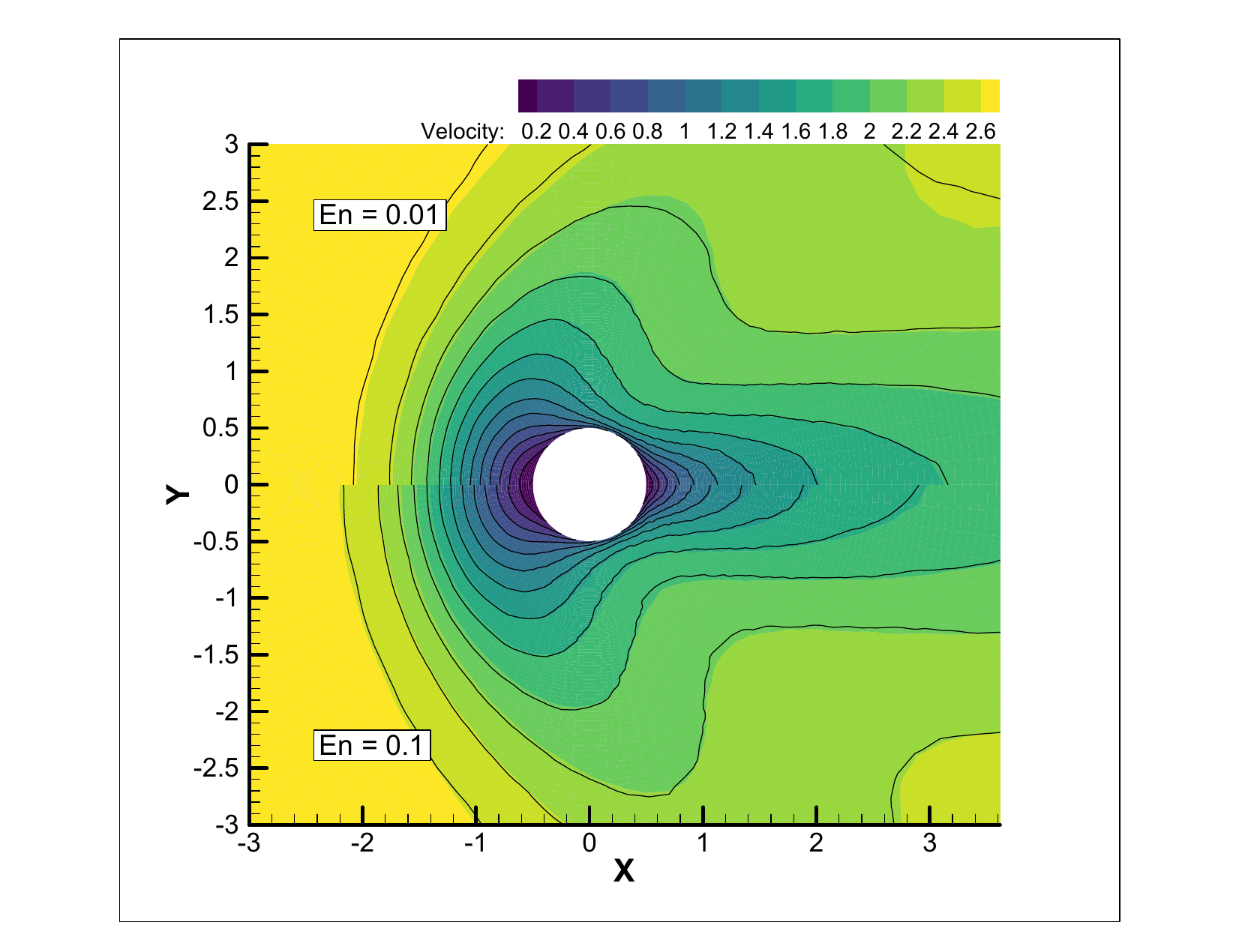}
\hfill
\includegraphics[width=0.32\textwidth,trim=100pt 40pt 100pt 40pt,clip]{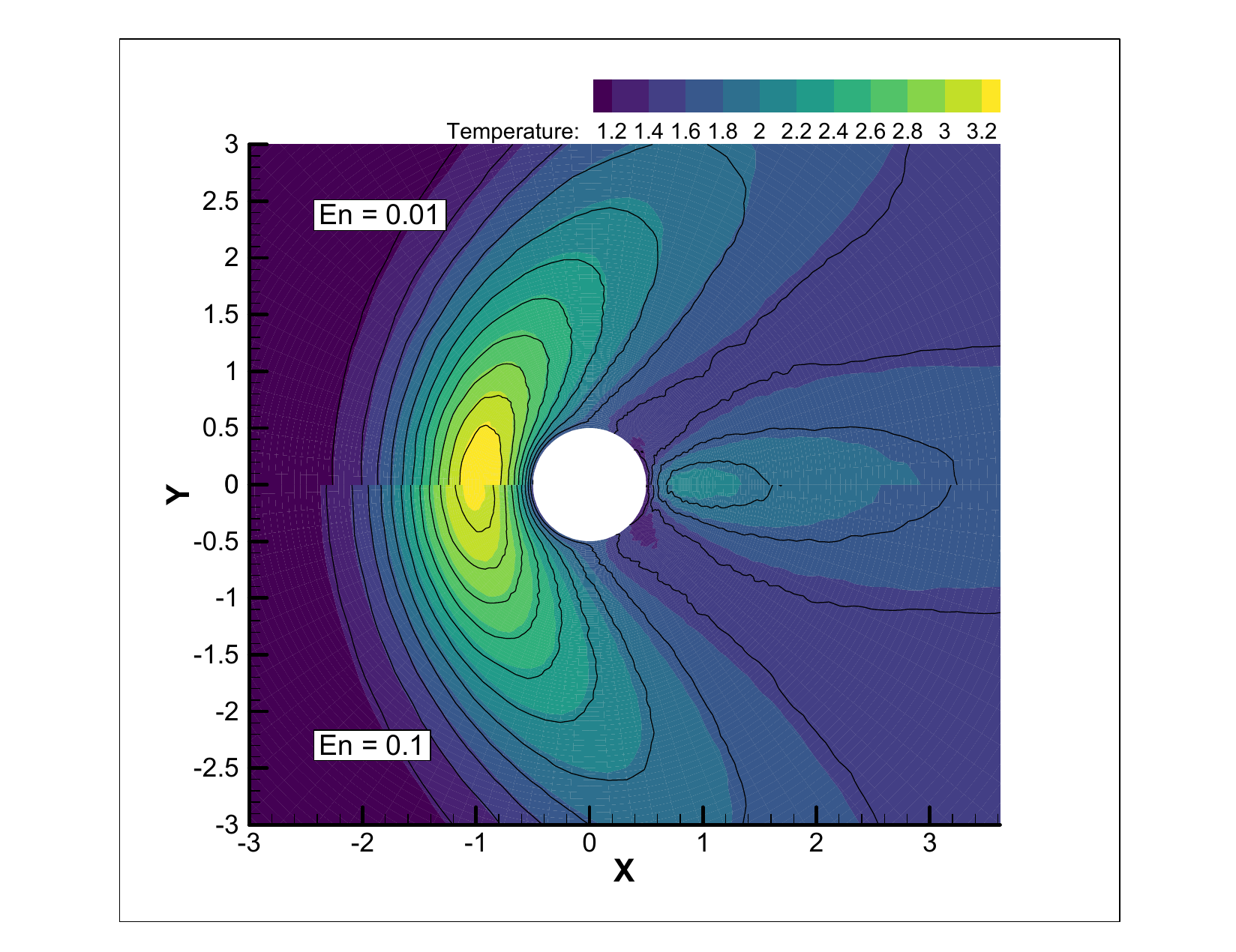}
\caption{First row: Geometry and meshes for the flow passing through a cylinder. Contours of  density, velocity, and temperatures at Ma = 3, (second row) Kn = 0.5 and (third row) Kn = 0.05. DIG and ESMC results for the Enskog equation are shown as solid black lines and the colored background, respectively.}
\label{fig:Contour_Ma3_cylinder_macro}
\end{figure}

\subsection{Supersonic flow passing cylinder}

\begin{figure}[t!]
    \centering
    \includegraphics[width=0.42\textwidth,trim = 100pt 40pt 100pt 40pt,clip]{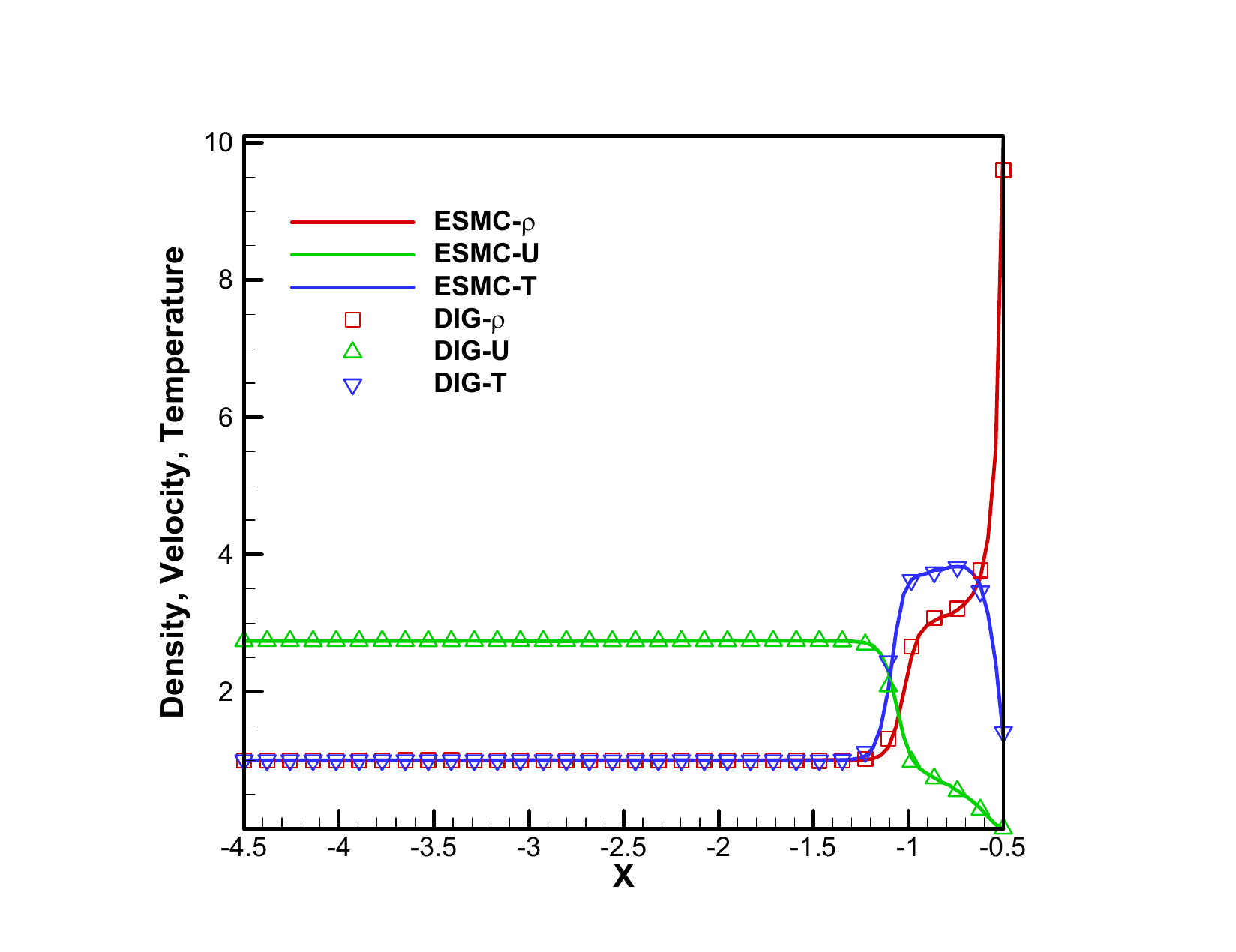}
    \includegraphics[width=0.42\textwidth,trim = 100pt 40pt 100pt 40pt,clip]{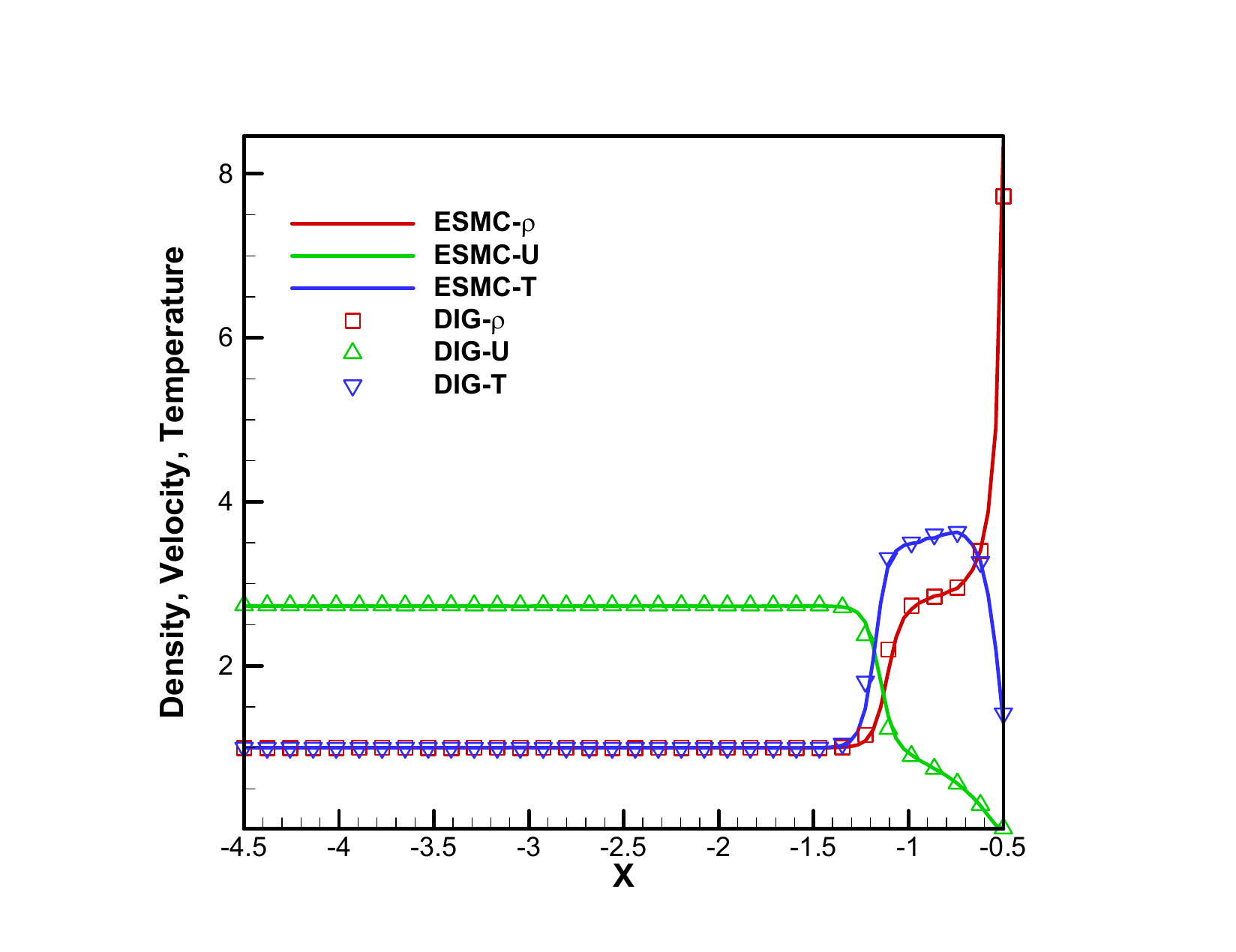}
    \\
    \includegraphics[width=0.42\textwidth,trim = 100pt 40pt 100pt 40pt,clip]{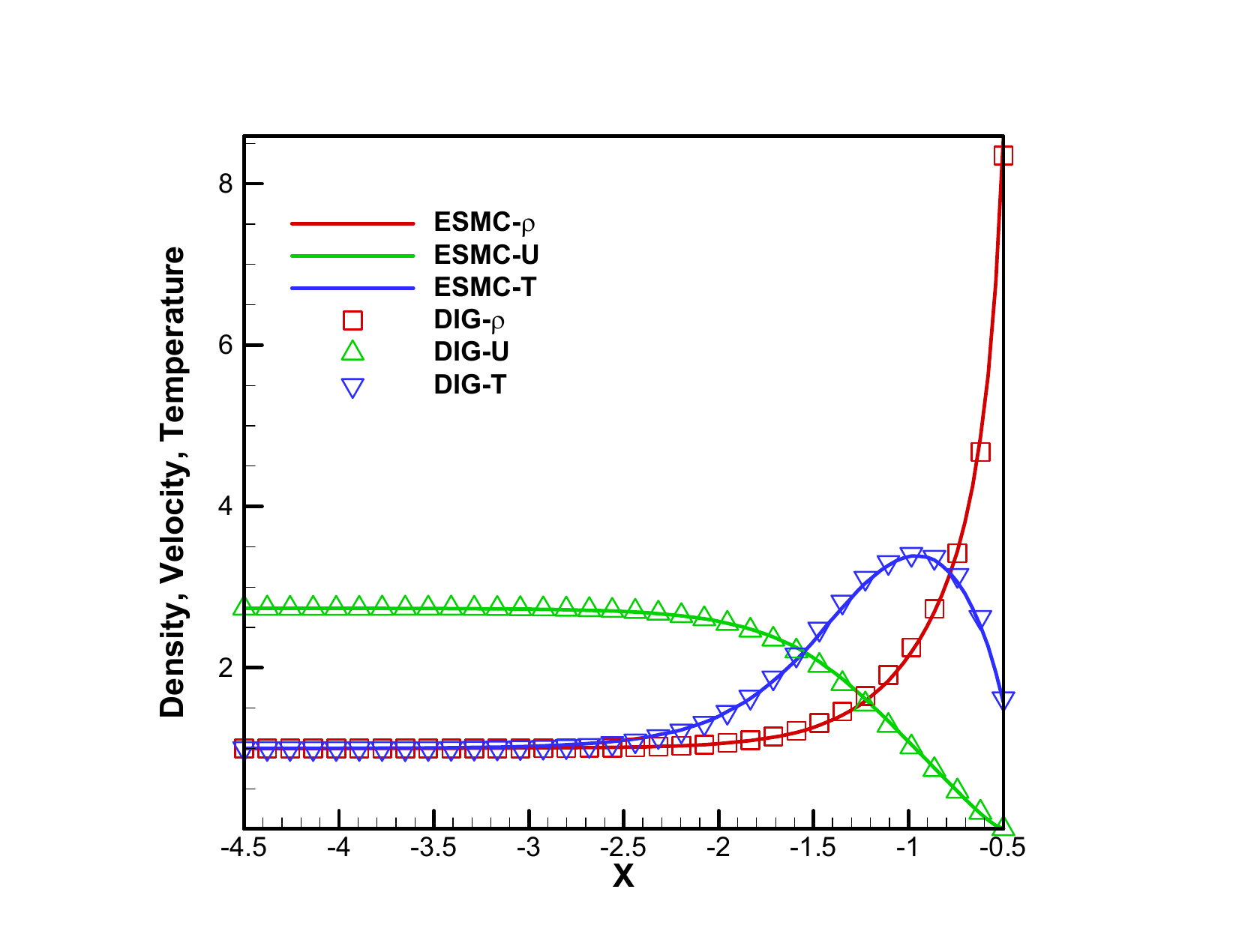}
    \includegraphics[width=0.42\textwidth,trim = 100pt 40pt 100pt 40pt,clip]{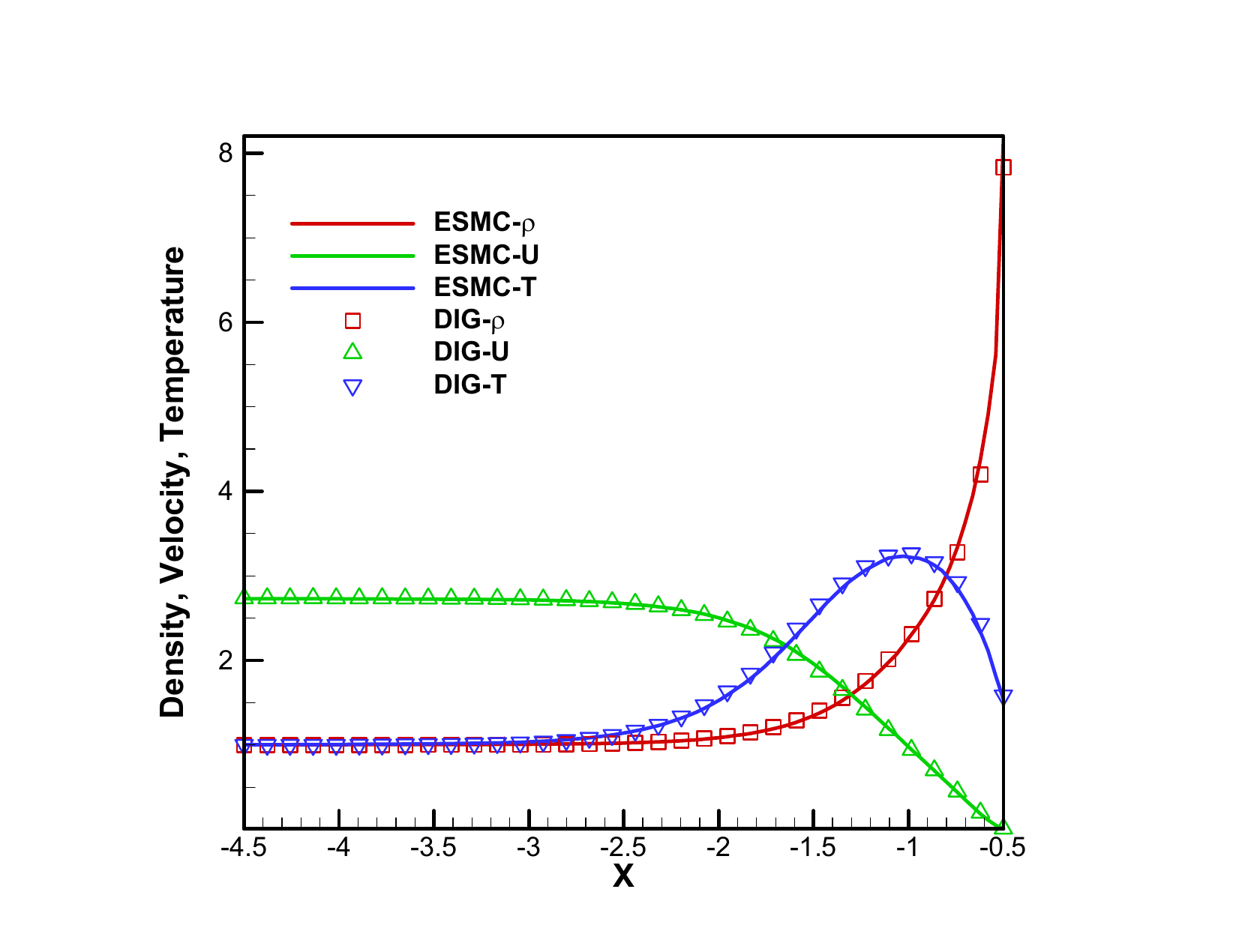}
    \caption{Comparisons of density, velocity and temperature along the stagnation line, when Ma = 3. The Knudsen numbers in the first and second rows are Kn = 0.05 and  0.5, respectively. The Enskog numbers in the first and second columns are En = 0.01 and 0.1, respectively.}
    \label{fig: stagnationline_macro}
\end{figure}

Consider the supersonic flow passing over a circular cylinder at $\text{Ma}_{\infty}=3$, where the Knudsen numbers vary from Kn = 0.05 to Kn = 0.5. The freestream number density $n_0$, temperature $T_0$, and the diameter of the cylinder $L_0$ are chosen to be the reference values. The computational domain is an annular region,  with an outer boundary of diameter $11L_0$ representing equilibrium free stream flow, and an inner boundary corresponding to the cylinder surface, maintained at a temperature of $T_w=T_0$, see the first row in Fig.~\ref{fig:Contour_Ma3_cylinder_macro}. The entire domain is discretized into $M \times N$ structured quadrilateral meshes, with refinement near the cylinder surface. Here, $M$ and $N$ denote the number of segments in the circumferential and radial directions, respectively. When Kn = 0.5, both ESMC and DIG employ a grid with $M=100$ and $N = 128$.  The thickness of the first cell layer adjacent to the cylinder surface is set to $0.01L_0$. When  Kn = 0.05, the grid is refined to $M=200$ and $N = 256$ in ESMC and $M=100$ and $N = 128$ in DIG.  An average of 100 particles is initialized in each cell, and particle velocities are sampled from an equilibrium distribution function with the same density and temperature as the freestream but zero initial velocity.

% It can be seen that the DIG results are consistent with the ESMC across the entire Enskog and Knudsen number domain. 

Figure~\ref{fig:Contour_Ma3_cylinder_macro} compares the contours of density, velocity, and temperature between the DIG and ESMC results, while 
Fig.~\ref{fig: stagnationline_macro} further compares the density, velocity, and temperature profiles along the stagnation line. Both ESMC and DIG show generally good agreement under different rarefaction and denseness conditions. The shock thickness and peak density of the cylinder wall surface decrease with the Knudsen number, but increase with the Enskog number. The latter arises because both the true shear viscosity~\eqref{mu_kap} and the bulk viscosity increase with En.

\begin{figure}[t]
    \centering
    \includegraphics[width=0.42\textwidth,trim = 100pt 40pt 100pt 40pt,clip]{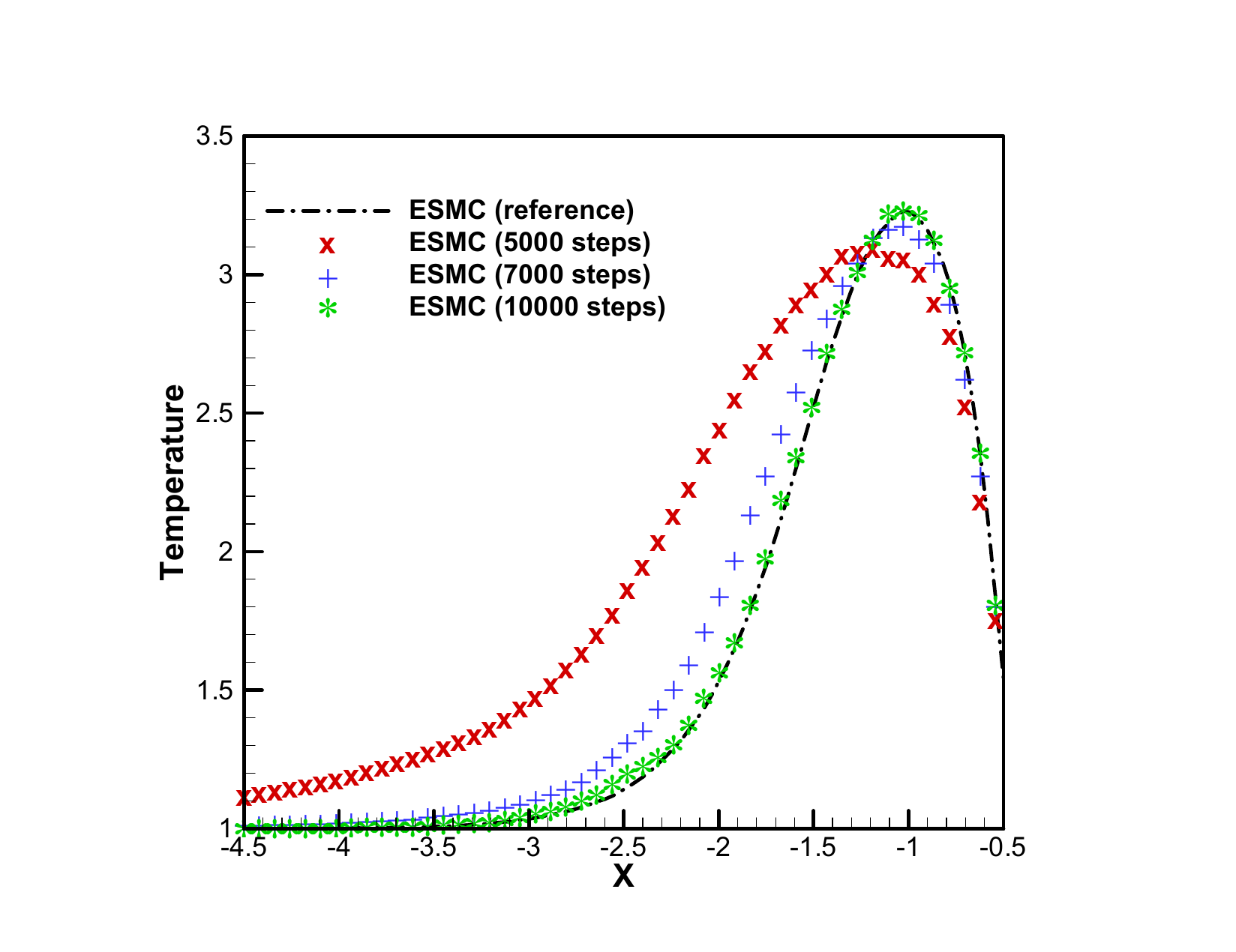}
    \includegraphics[width=0.42\textwidth,trim = 100pt 40pt 100pt 40pt,clip]{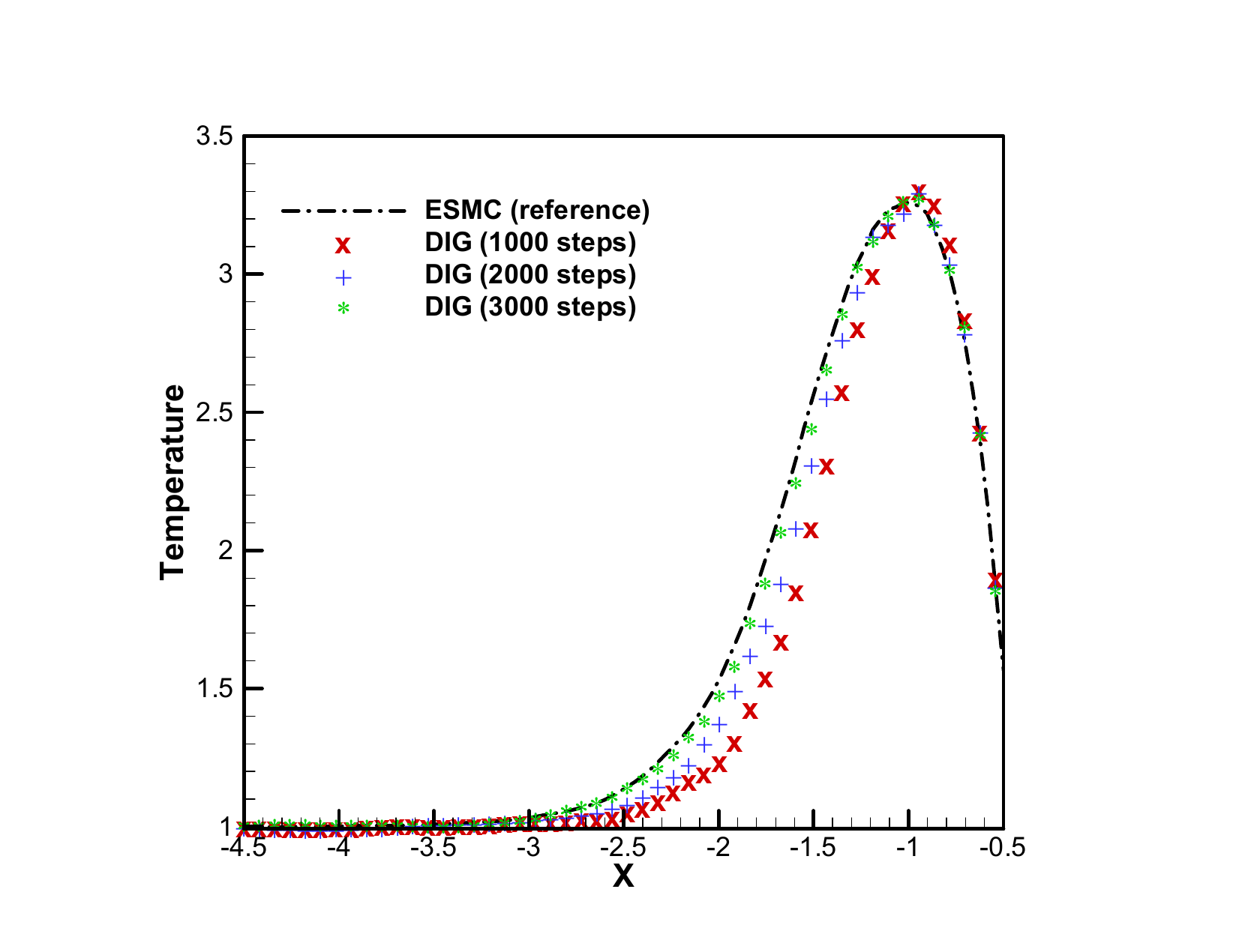}
    \\
    \includegraphics[width=0.42\textwidth,trim = 100pt 40pt 100pt 40pt,clip]{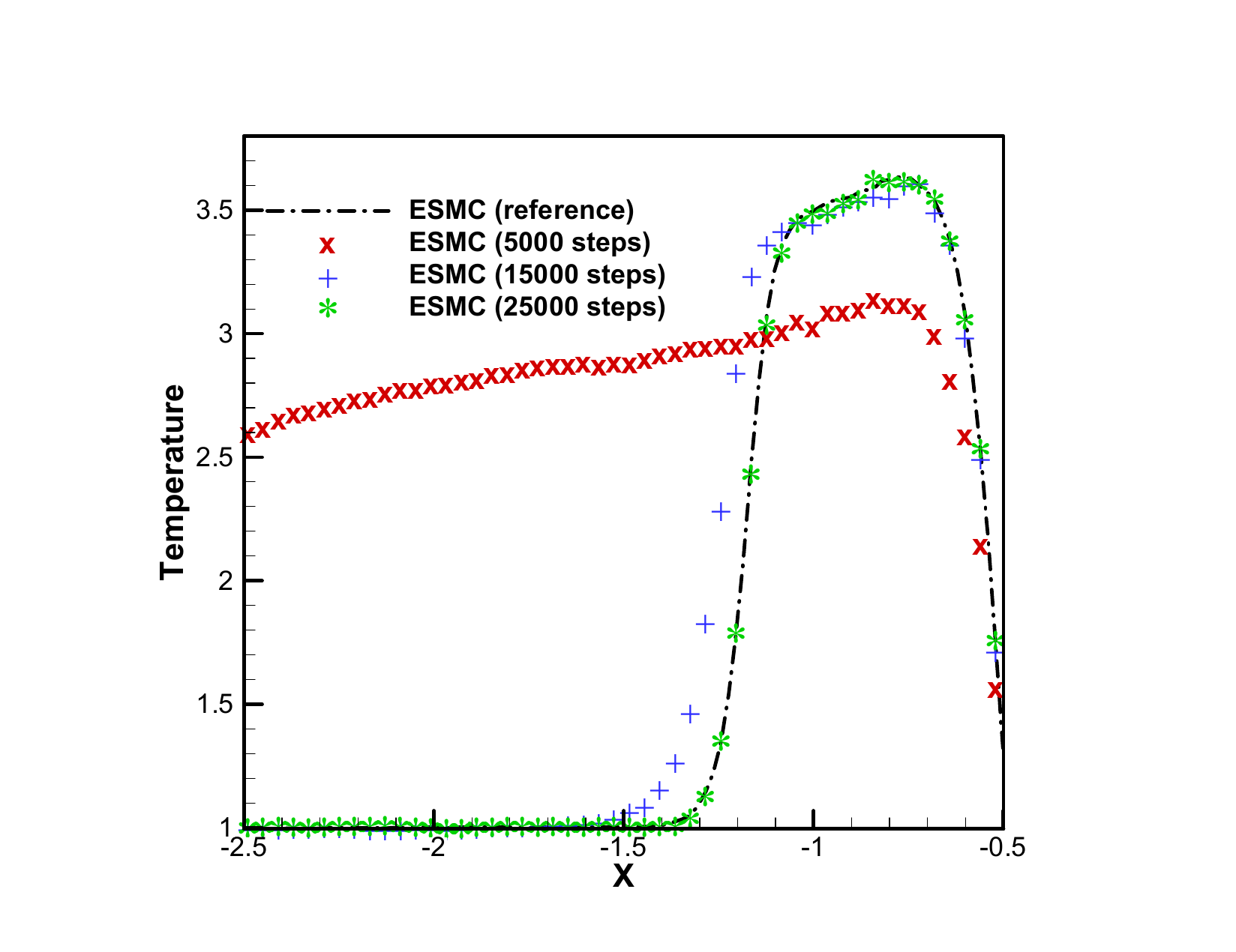}
    \includegraphics[width=0.42\textwidth,trim = 100pt 40pt 100pt 40pt,clip]{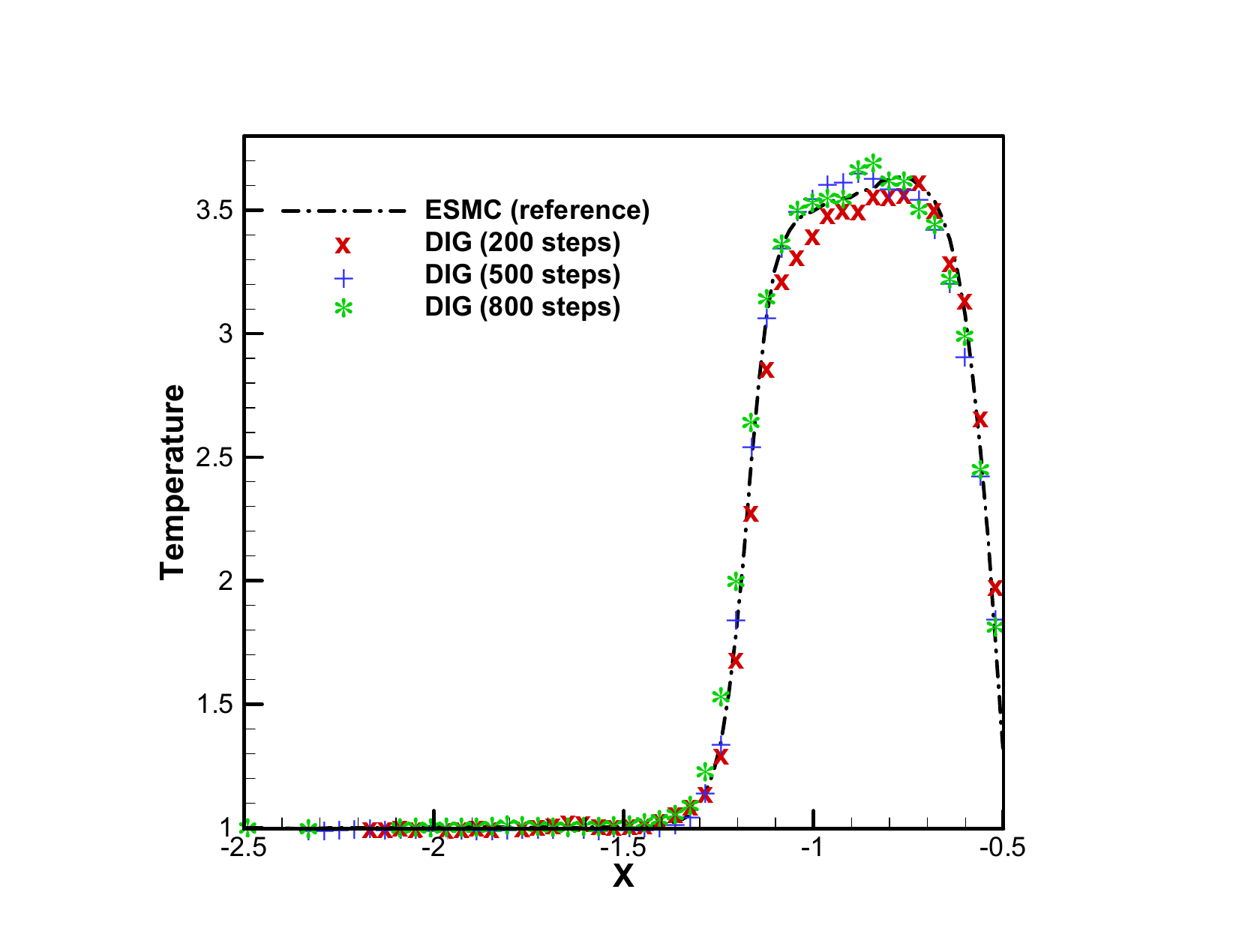}
    \caption{Convergence history of the temperature along the stagnation line, when Kn = 0.5 (first row) and Kn = 0.05 (second row) at En = 0.1.}
    \label{fig: stagnationline_evolution_T}
\end{figure}

Figure \ref{fig: stagnationline_evolution_T} illustrates the convergence history of temperature along the stagnation line. When Kn = 0.5 and En = 0.1, ESMC requires approximately 10000 iterations to reach the steady state, while DIG converges within 3000 iterations, achieving equivalent accuracy with approximately one-third of the computational steps. When Kn decreases to 0.05, the ESMC requires 25000 iterations, whereas DIG reaches steady state within only 800 steps. The acceleration to steady state achieved by DIG become even more pronounced, with the required number of steps reduced by nearly a factor of 30 compared with the ESMC.

The computational time is summarized in Table~\ref{tab:dsmc_dig_comparison}. It should be noted that the macroscopic synthetic equations \eqref{macro} are solved every 100 ESMC steps, i.e., $N_m = 100$ in Fig.~\ref{fig:DIG_flowchart}. At Kn = 0.5, DIG reduces the total simulation time by roughly one-third compared to ESMC. At Kn = 0.05, the advantage of DIG becomes more pronounced. The iterative solution of the macroscopic equations modifies the particle distribution evolution, thereby accelerating convergence to the steady-state solution. Moreover, incorporating the macroscopic synthetic equations grants DIG the asymptotic-preserving property, so that less spatial grids can be used. As a result, when Kn=0.05, DIG reduces the CPU time by up to two orders of magnitude relative to ESMC, underscoring its superior efficiency. The computational savings will become even more significant further down the near-continuum regime.

\begin{table}[t]
\centering
\caption{Comparison of ESMC and DIG under Kn and En, in the two-dimensional supersonic flow passing cylinder. The simulation time is reported as wall-clock time, measured in minutes. The ESMC employs OpenMP parallelization with 40 cores, whereas the NS solver runs on a single core.
}
\label{tab:dsmc_dig_comparison}
\begin{tabular}{ccccccccc}
\toprule
\multirow{2}{*}{Kn} & 
\multirow{2}{*}{En} & 
\multirow{2}{*}{method} & 
\multirow{2}{*}{$N_{\text{cell}}$} & 
\multicolumn{2}{c}{Transition state} & 
\multicolumn{2}{c}{Steady state} \\
\cmidrule(lr){5-6} \cmidrule(lr){7-8}
& & & & steps & time  & steps & time \\
\midrule
\multirow{2}{*}{0.5} & 0.1 & ESMC &  100$\times$128 &10000 & 18 & 10000 & 16 \\
& & DIG & 100$\times$128 & 3000 & 6 & 3000 & 5 \\
\cmidrule(lr){2-8}
& 0.01 & ESMC & 100$\times$128 & 7000 & 11 & 10000 & 15 \\
& & DIG & 100$\times$128 & 1000 & 2 & 3000 & 5 \\
\midrule
\multirow{2}{*}{0.05} & 0.1 & ESMC & 200$\times$256 & 30000 & 505 & 10000 & 156 \\
& & DIG & 100$\times$128 & 1000 & 3 & 3000 & 11 \\
\cmidrule(lr){2-8}
& 0.01 & ESMC & 200$\times$256 & 20000 & 341 & 10000   & 147 \\
& & DIG &  100$\times$128 & 500 & 2 & 3000 & 11 \\
\bottomrule
\end{tabular}
\end{table}

\subsection{Dense flow in porous media}

\begin{figure}[!t]
        \centering
       \hspace{0.9cm} \includegraphics[width=0.45\linewidth, trim=80pt 30pt 100pt 85pt,clip ]{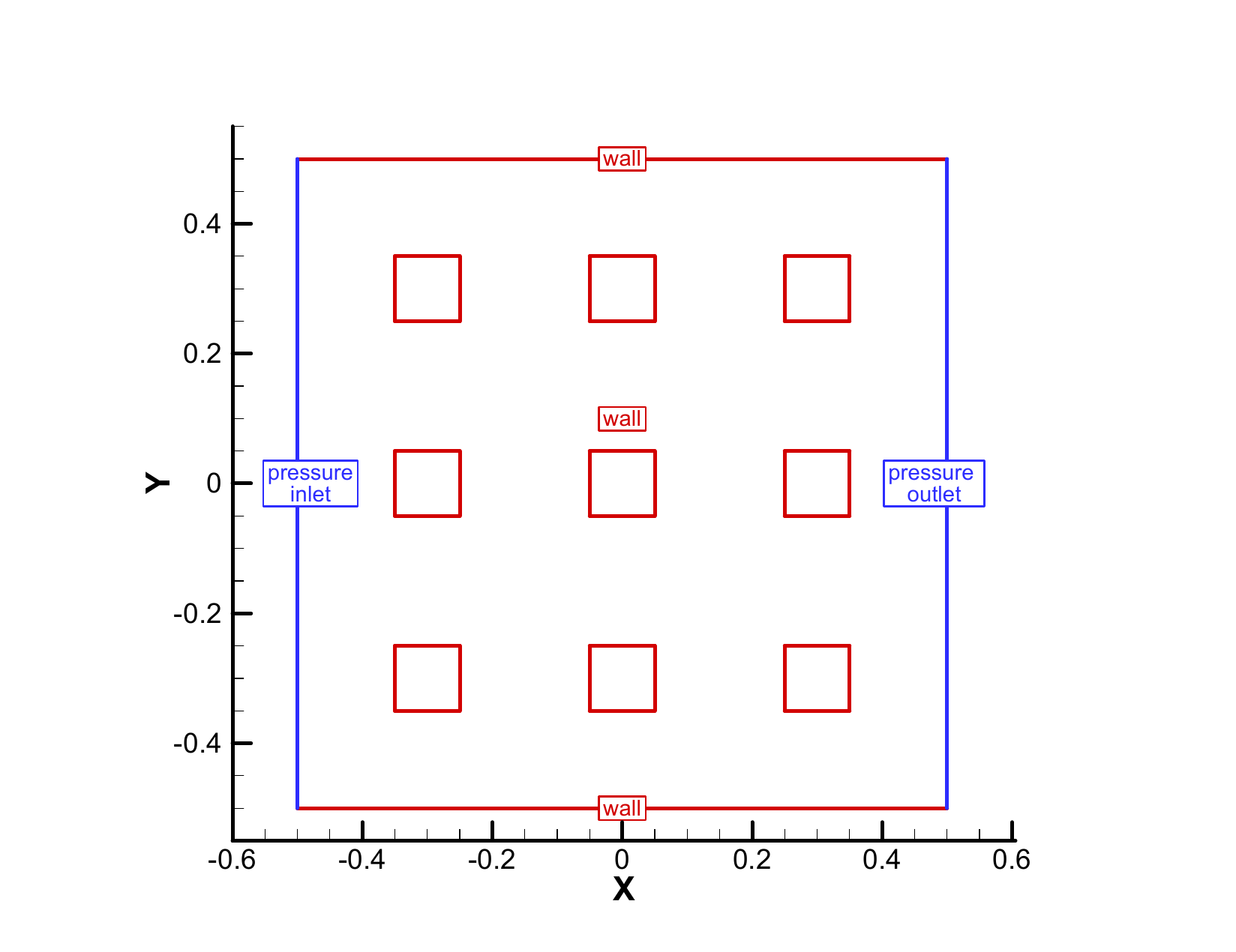}
\includegraphics[width=0.45\linewidth, trim=20pt 40pt 150pt 85pt, clip]{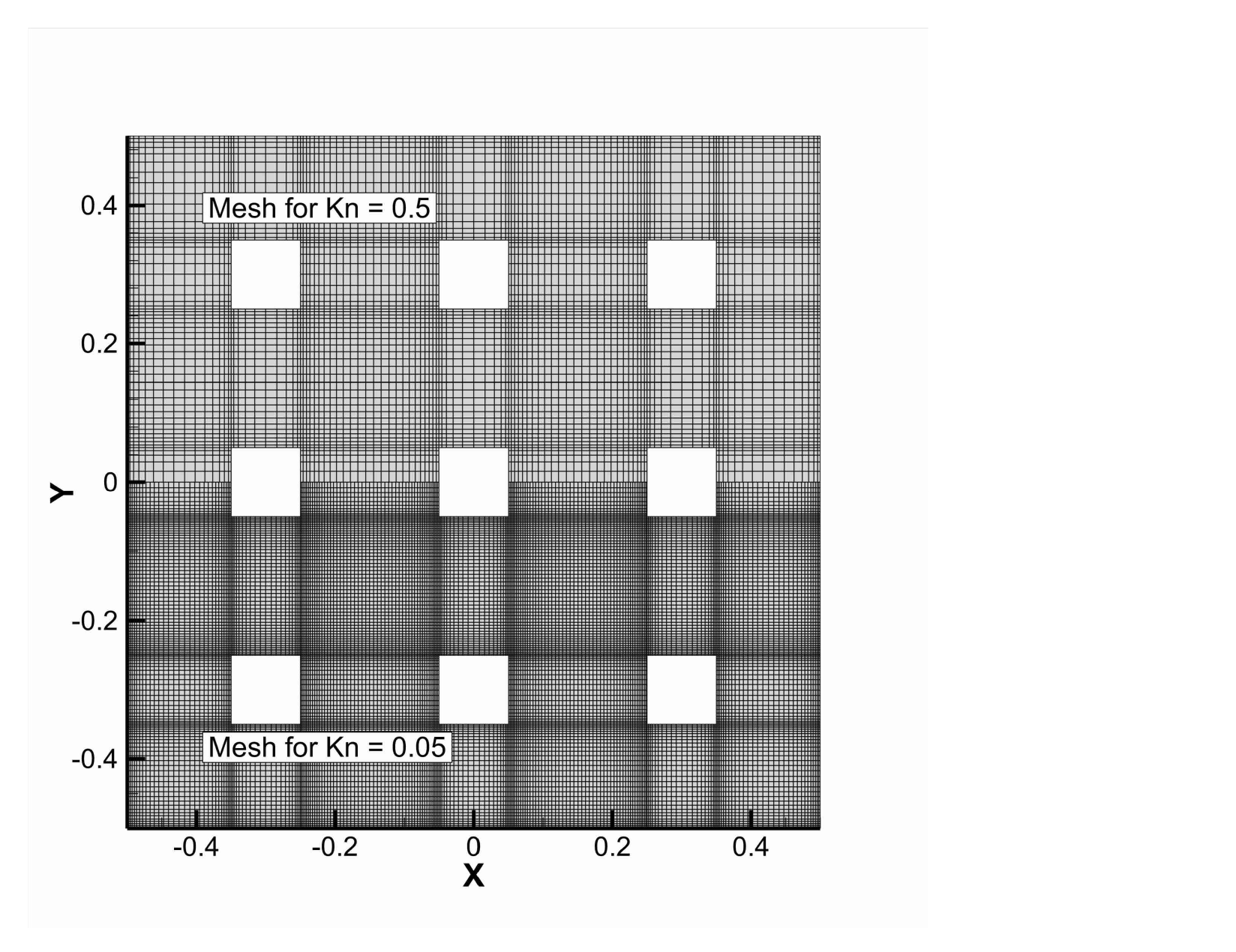}\\
        \vspace{0.2cm}
\includegraphics[width=0.48\textwidth,trim=30pt 40pt 250pt 40pt,clip]{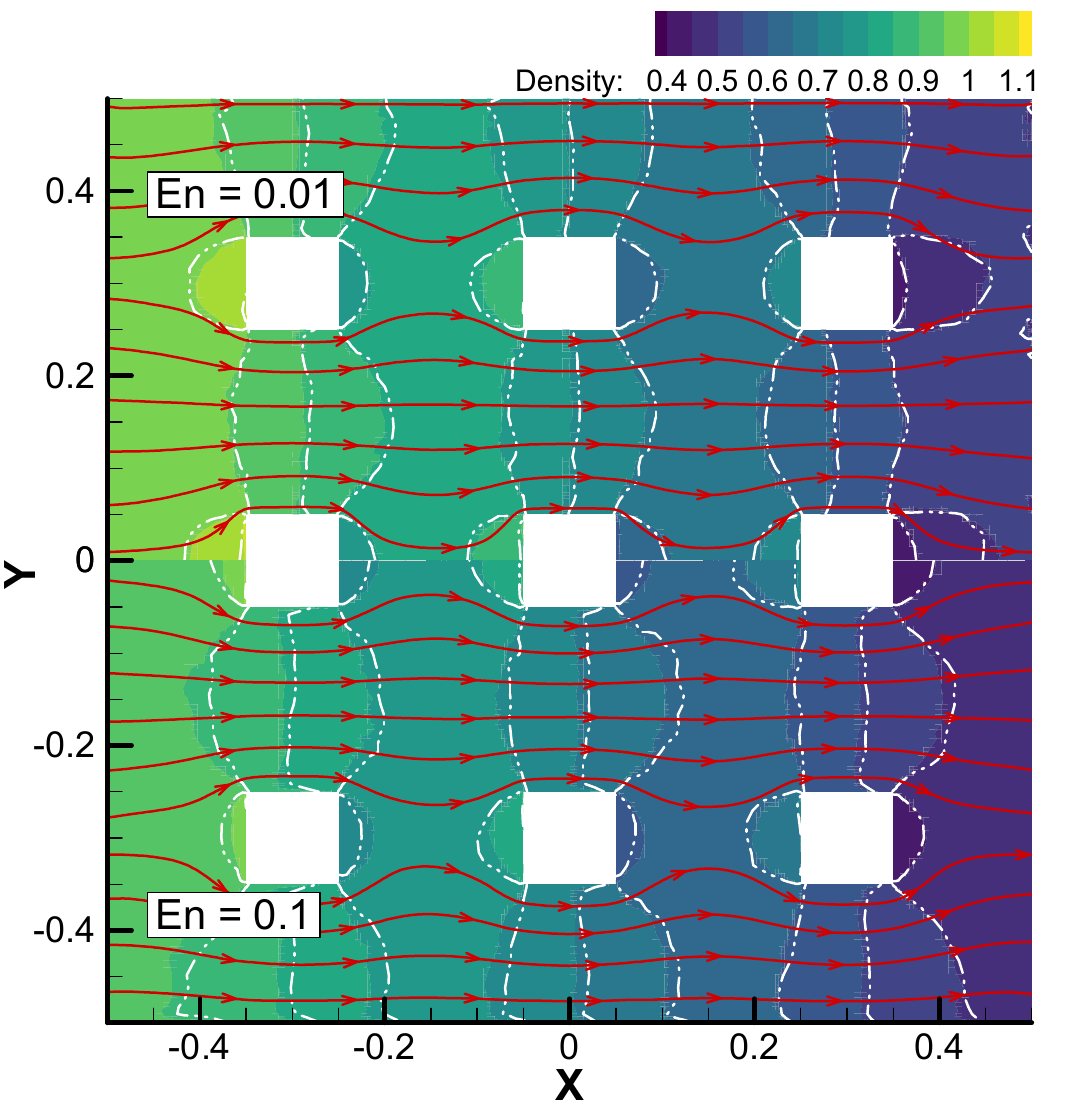}     \includegraphics[width=0.48\textwidth,trim=30pt 40pt 250pt 40pt,clip]{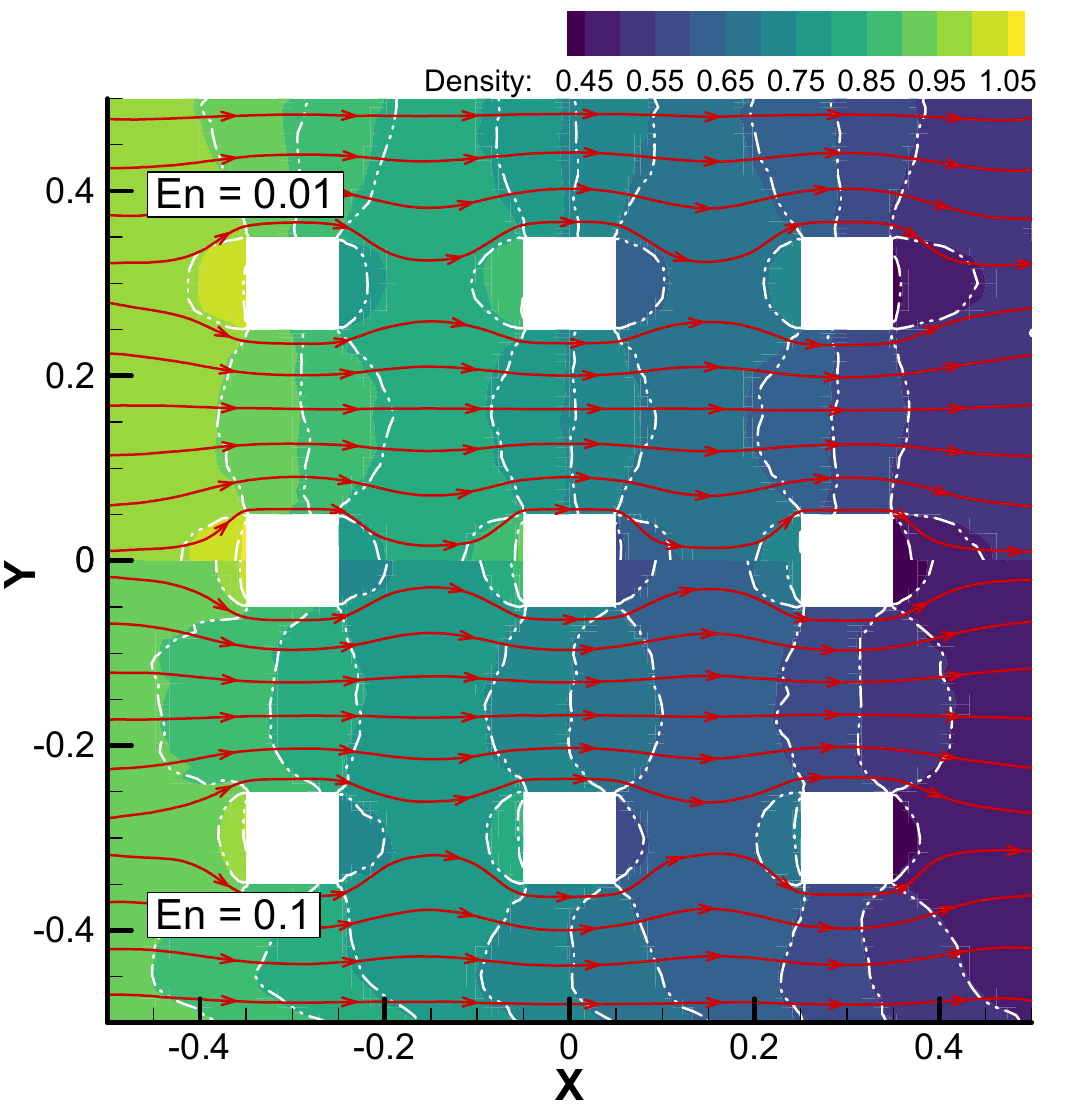}   
        \\
        \vspace{0.2cm}
\includegraphics[width=0.48\textwidth,trim=30pt 260pt 250pt 40pt,clip]{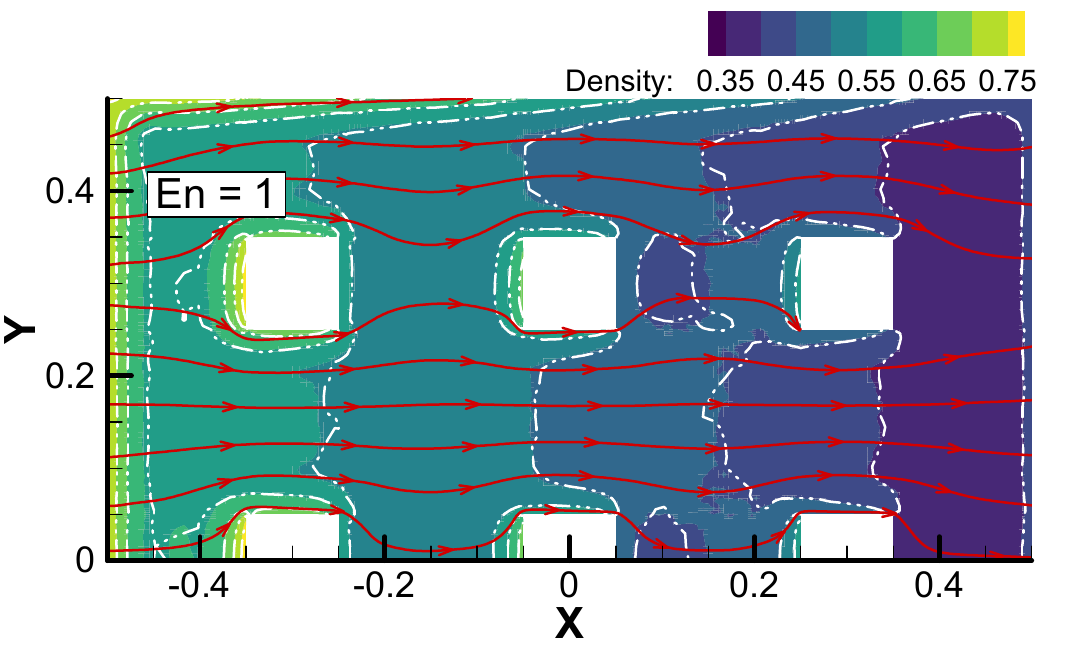}
\includegraphics[width=0.48\textwidth,trim=30pt 260pt 250pt 40pt,clip]{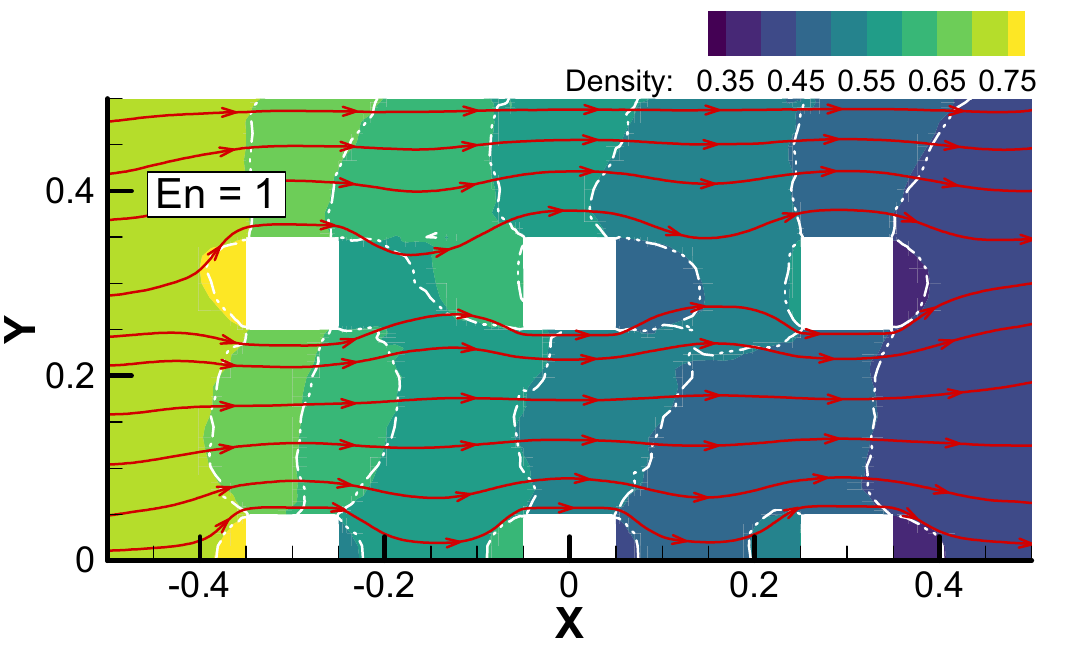}
    \caption{First row: Geometry and meshes for porous media. Other rows: Density contours (DIG and ESMC results are shown as dashed white lines and the colored background, respectively) and the velocity streamlines (red arrow lines) at  Kn = 0.05 (left column) and  Kn = 0.5 (right column).} 
    \label{fig:streamline_rho_porous}%trim = left bottom right top
\end{figure}

We simulate the rarefied flows of a dense gas in porous media, mimicking the shale gas extraction. The channel height $H$ is the characteristic flow length. This domain is discretized using a non-uniform Cartesian grid featuring mesh refinement near the surface to capture the Knudsen layer and adsorption layer, see Fig.~\ref{fig:streamline_rho_porous}. The number of segments in the $x$ and $y$ directions is shown in Table~\ref{tab:esmc_dig_comparison_porous}.

At the left boundary, a constant temperature $T_{\text{in}}=1$ and pressure $p_{\text{in}}=P_0$ is applied, where the density $n_{\text{in}}$ is determined by the equation of state: $p_{\text{in}}=n_{\text{in}}k_BT_{\text{in}}(1+b n_{\text{in}}\chi_{\text{in}})$. At the right boundary,  the pressure is maintained at $p_{\text{out}}=n_{\text{out}}k_BT_{\text{out}}(1+b n_{\text{out}}\chi_{\text{out}})=0.5P_0$.
To promptly convey the pressure information from dense gas flows at both inlet and outlet, the macroscopic quantities on the inner side of the interface are evaluated as follows~\cite{gsis_Dense}:
\begin{equation}
    \begin{aligned}
        &u_{\text{in},x}= u_{i,x} + \frac{p_{\text{in}}-p_i}{mn_ia_i},\quad u_{\text{in},y}=0,\quad
        n_{\text{out}}=n_i + \frac{p_{\text{out}}-p_i}{ma_i^2},\\
        &u_{\text{out},x}=u_{i,x} + \frac{p_i-p_{\text{out}}}{mn_ia_i},\quad u_{\text{out},y}= u_{i,y}, \quad T_{\text{out}}= \frac{p_{\text{out}}}{n_{\text{out}}k_B(1+bn_{\text{out}}\chi_{\text{out}})},
\end{aligned}\label{inlet_outlet_condition}
\end{equation}
where the subscript $i$ represent the local macroscopic quantities, $a_i =\sqrt{\gamma R T_i}$ denotes the local speed of sound speed, and $p_i=n_ik_BT_i(1+bn_i\chi_i)$ represents the local pressure.

\begin{figure}[p]
    \centering
    \includegraphics[width=0.42\textwidth,trim = 20pt 60pt 250pt 60pt,clip]{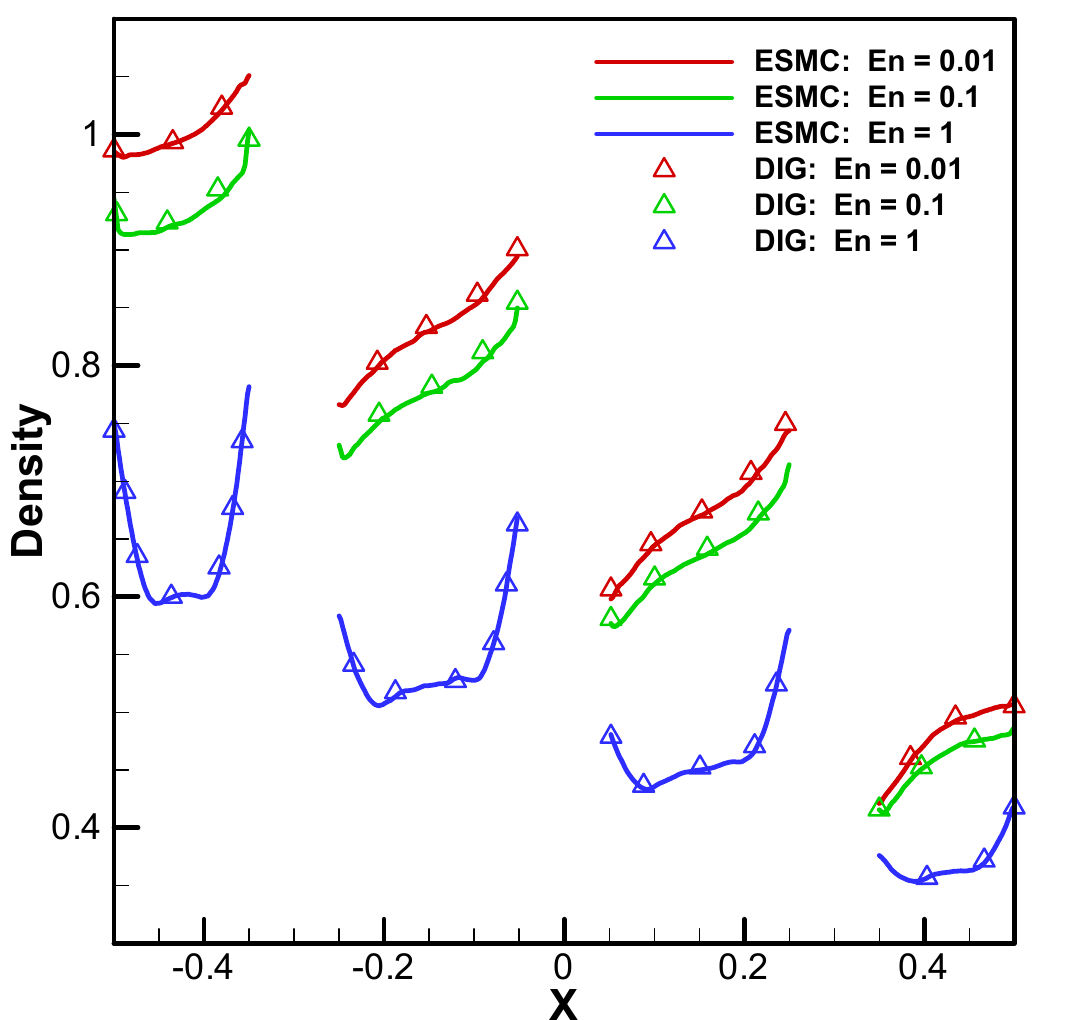}
    \includegraphics[width=0.42\textwidth,trim = 20pt 60pt 250pt 60pt,clip]{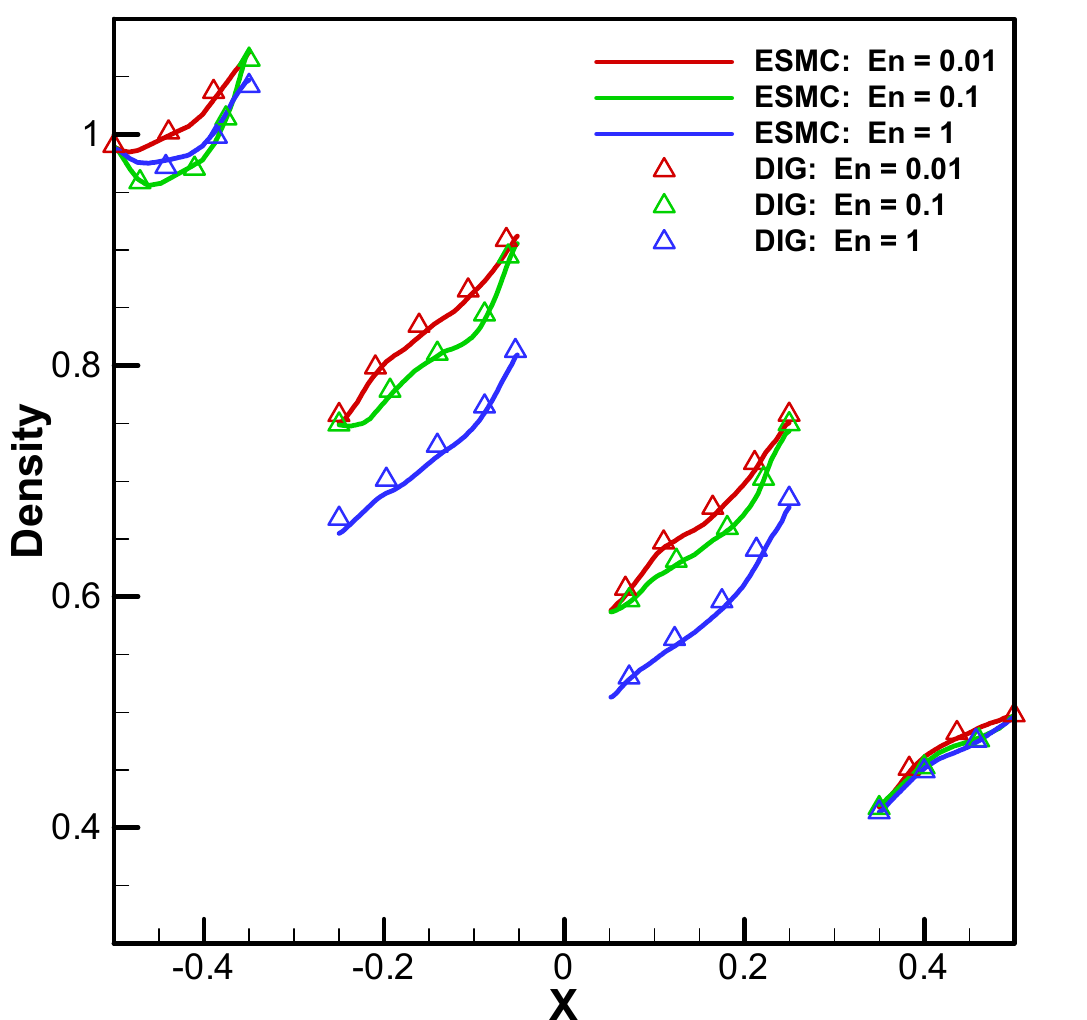}
    \\
      \includegraphics[width=0.42\textwidth,trim = 20pt 40pt 250pt 60pt,clip]{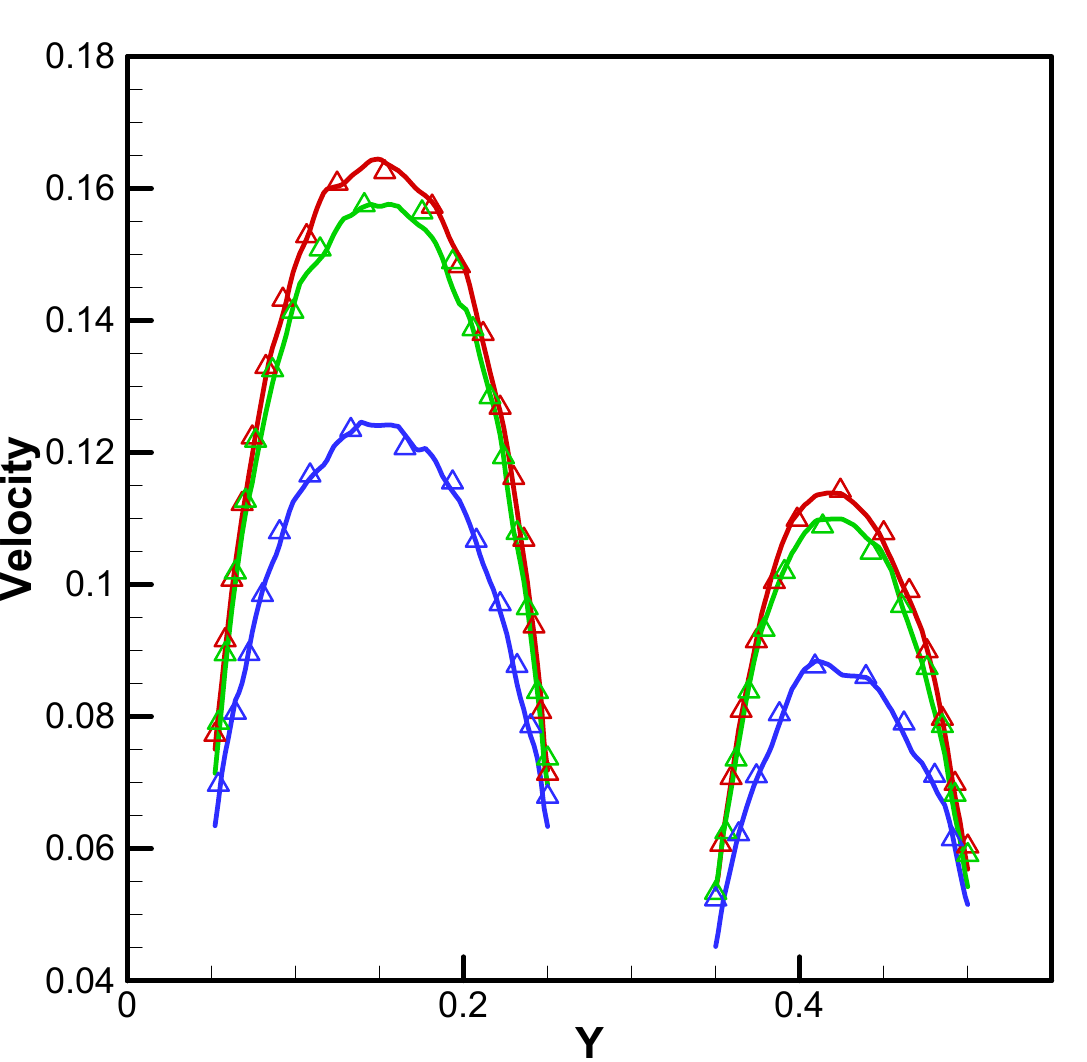}
        \includegraphics[width=0.42\textwidth,trim = 20pt 40pt 250pt 60pt,clip]{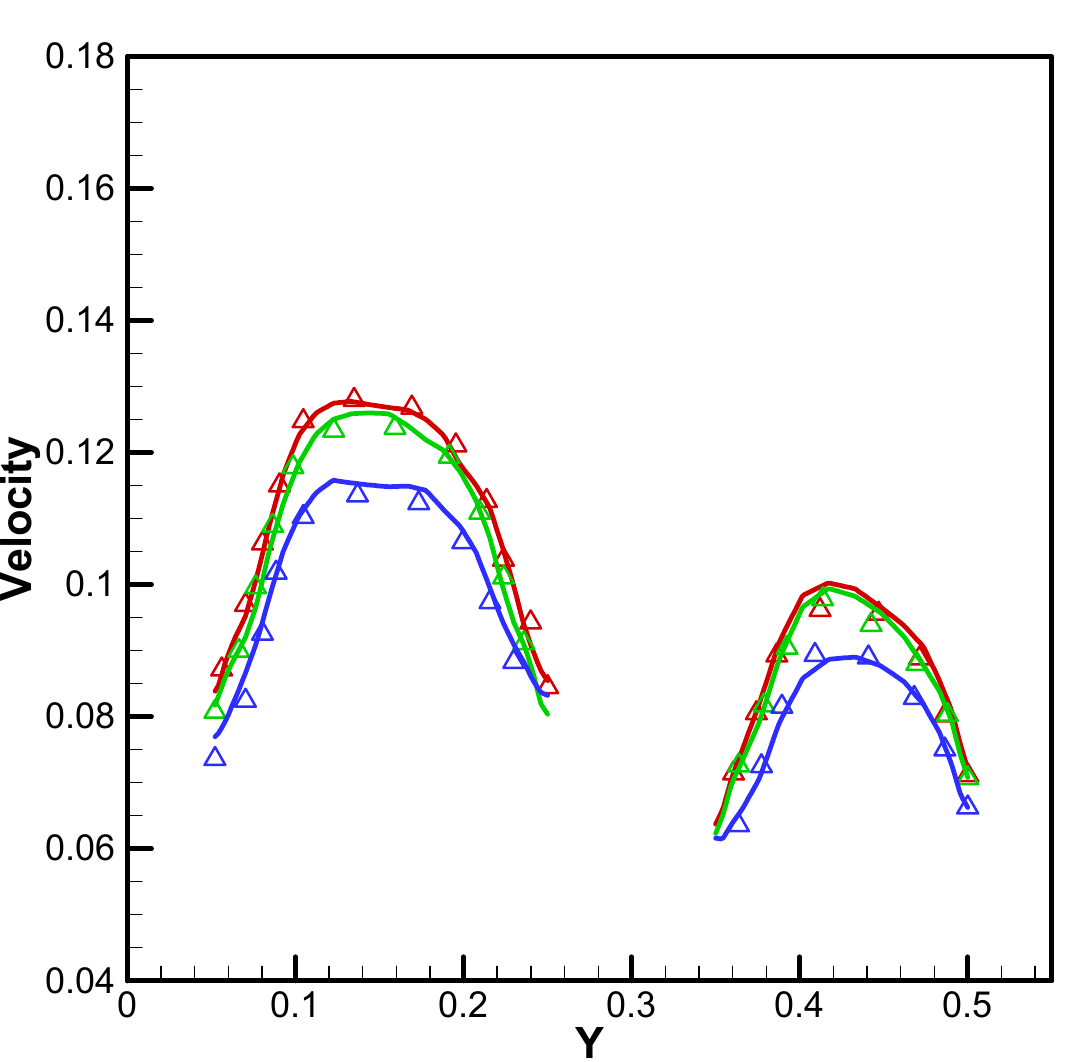}
    \\
    \includegraphics[width=0.42\textwidth,trim = 20pt 40pt 250pt 60pt,clip]{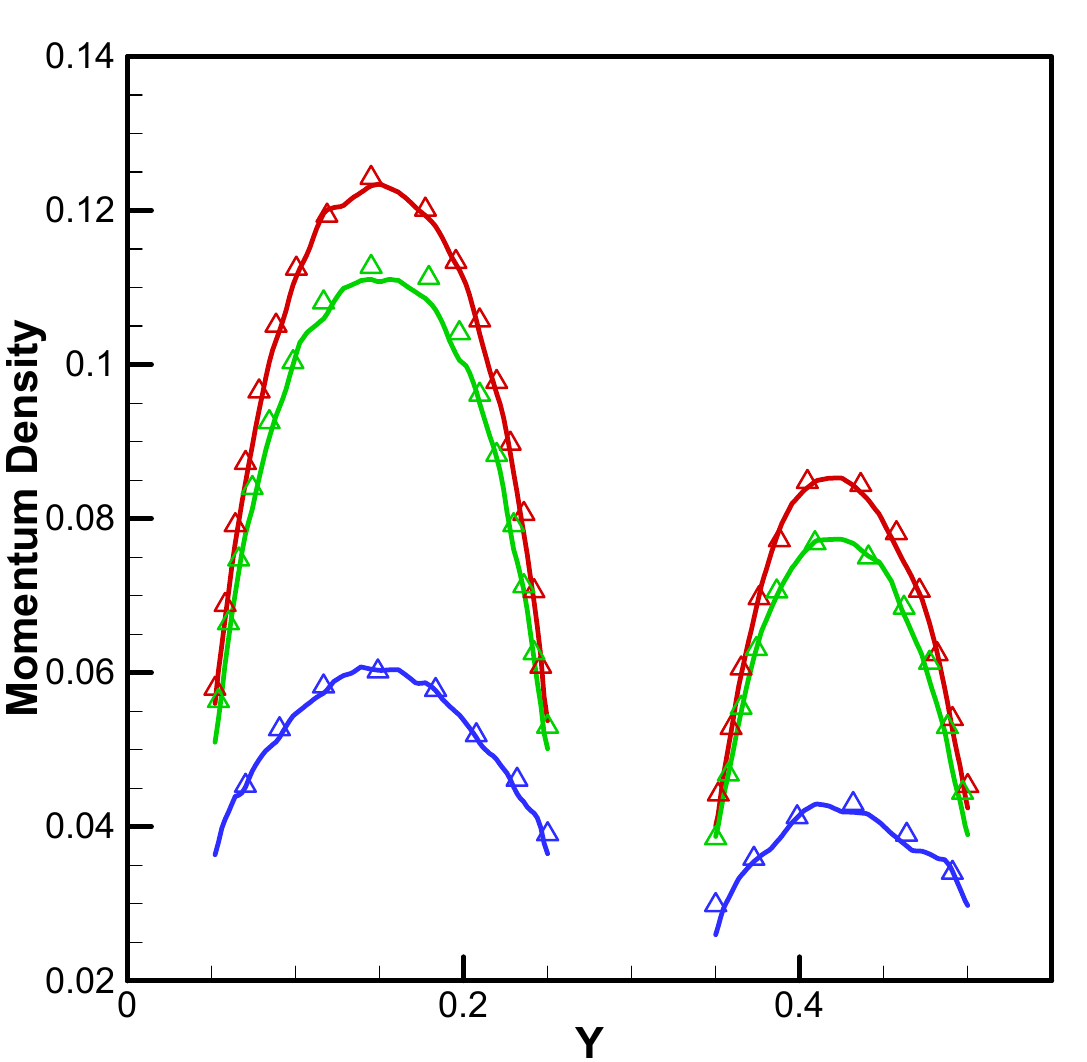}
    \includegraphics[width=0.42\textwidth,trim = 20pt 40pt 250pt 60pt,clip]{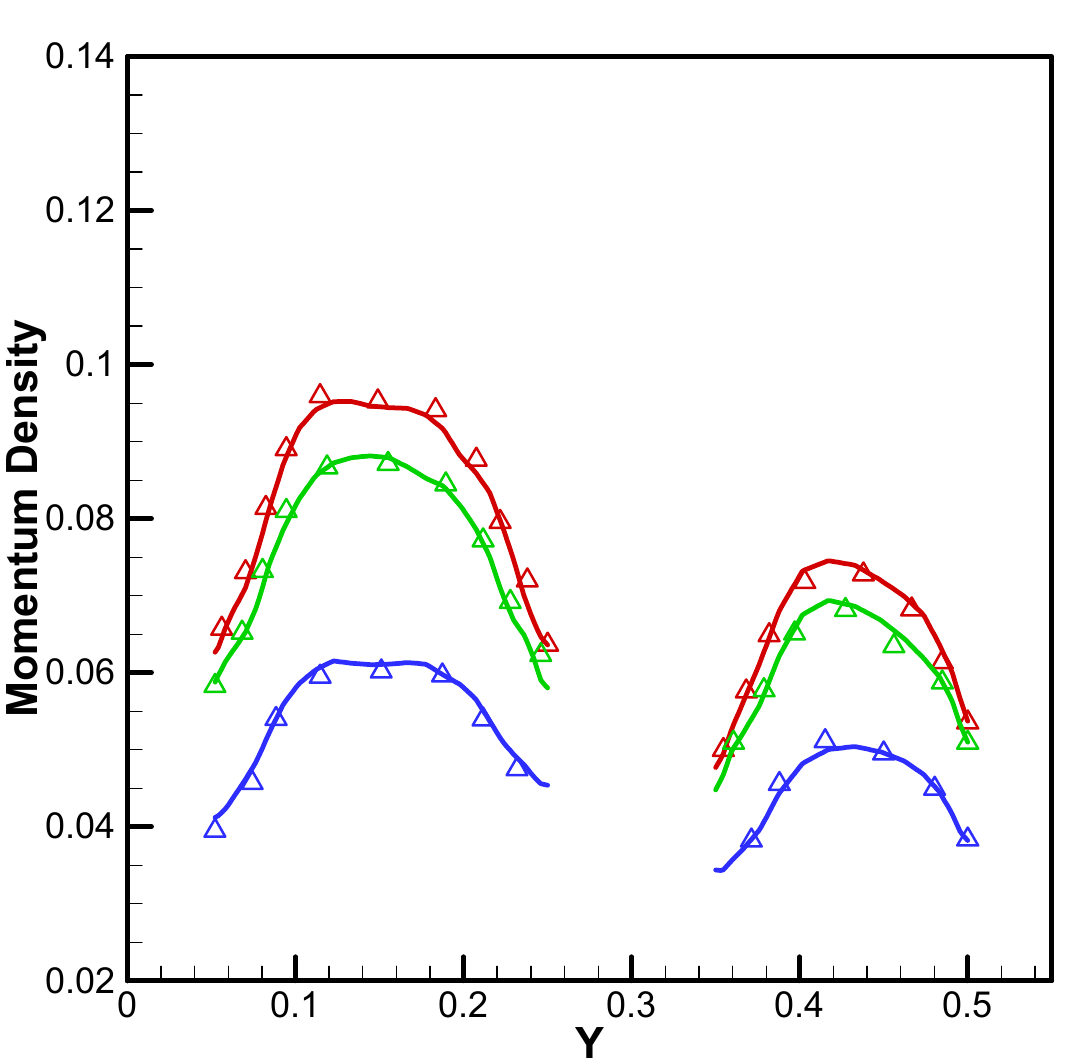}
    \caption{The density $\rho$ at $y=0$ (first row), velocity $u_x$ at $x=0$ (second row), and momentum density $\rho u_x$ at $x=0$ (third row). The Knudsen numbers in the left and right columns are 0.05 and 0.5, respectively.}
    \label{fig:rho_mv_DIG_ESMC_compare}
\end{figure}

Figure~\ref{fig:streamline_rho_porous} shows that the density contours obtained from the DIG and ESMC agree well with each other. In the dilute gas limit, the contours are nearly vertical within the region occupied by the solid blocks. However, as En increases while the pressure drop remains fixed, the density decreases due to the non-ideal equation of state. Moreover, the density contours exhibit strong distortions, clearly indicating the formation of pronounced adsorption layers near the solid blocks, as shown in Fig.~\ref{fig:rho_mv_DIG_ESMC_compare}. When the particles flow from the left to right due to the pressure difference, the density gradually decreases along the flow direction without encountering the blocks, whereas the presence of the blocks causes the density to increase from left to right between blocks. However, when En is large, the adsorption layer is appreciable, which causes a large number of particles to gather near these blocks, resulting in a significant increase in density near the solid blocks.

\begin{table}[t!]
\caption{Comparisons of ESMC and DIG in porous media flow simulations. The simulation time is reported as wall-clock time, measured in minutes. The ESMC employs OpenMP parallelization with 40 cores, whereas the NS solver runs on a single core.
}
\centering
\begin{tabular}{ccccccccc}
\toprule
\multirow{2}{*}{Kn} & 
\multirow{2}{*}{En} & 
\multirow{2}{*}{method} & 
\multirow{2}{*}{$N_{\text{cell}}$} & 
\multicolumn{2}{c}{Transition state} & 
\multicolumn{2}{c}{Steady state} \\
\cmidrule(lr){5-6} \cmidrule(lr){7-8}
& & & & steps & time  & steps & time \\
\midrule
\multirow{3}{*}{0.5} & 0.01 & ESMC &  110$\times$110 & 5000 & 9 & 10000 & 16 \\
& & DIG & 110$\times$110 & 2000 & 4 & 3000 & 6 \\
\cmidrule(lr){2-8}
& 0.1 & ESMC & 110$\times$110 & 5000 & 11 & 10000 & 26 \\
& & DIG & 110$\times$110 & 2000 & 4 & 3000 & 5 \\
\cmidrule(lr){2-8}
& 1 & ESMC & 110$\times$110 & 5000 & 6 & 10000 & 16 \\
& & DIG & 110$\times$110 & 3000 & 5 & 3000 & 6 \\
\midrule
\multirow{3}{*}{0.05} & 0.01 & ESMC &220$\times$220 & 10000 & 78 & 10000 & 74 \\
& & DIG & 110$\times$110 & 1000 & 3 & 3000 & 8 \\
\cmidrule(lr){2-8}
&  0.1 & ESMC &  220$\times$220  & 7000 & 56 & 10000  & 74 \\
& & DIG &  110$\times$110 & 1000 & 3 & 3000 & 9 \\
\cmidrule(lr){2-8}
& 1 & ESMC &  220$\times$220 & 7000 & 41 & 10000 & 53 \\
& & DIG & 110$\times$110 & 2000 & 4 & 3000 & 7 \\
\bottomrule
\end{tabular}
\label{tab:esmc_dig_comparison_porous}
\end{table}

Figure~\ref{fig:rho_mv_DIG_ESMC_compare} also compares the flow velocity and momentum density $\rho u_x$ at $x=0$ obtained by the ESMC and DIG methods. It is observed that for a fixed Kn, the velocity slip shows minor sensitivity to the Enskog number. In contrast, the peak velocity decreases noticeably as En increases, and, together with the reduction in density, this leads to a rapid decline in momentum density.
When the En is held constant, the peak velocity reduces with increasing Kn, while the velocity slip exhibits a slight rising trend. 
As a result, the momentum density decreases rapidly as 
Kn increases for small values of En. However, when En=1, the blue lines suggest that the mass flow rates for Kn=0.05 and 0.5 are nearly the same. These results are consistent with the one-dimensional force-driven Poiseuille flow solved by both the fast spectral method of the Enskog equation~\cite{wu2016non} and the event-driven molecular dynamics simulations~\cite{Sheng2020PoF}, where the Knudsen minimum might disappear.

Table~\ref{tab:esmc_dig_comparison_porous} compares the computational cost between ESMC and DIG. 
At Kn = 0.5 and En = 0.01, ESMC requires approximately 5000 iterations to reach the steady state on the same grid, while DIG converges within 2000 iterations, achieving comparable accuracy with half the number of iteration steps and computational time. When Kn reduces to 0.05, ESMC necessitates 10000 iterations, whereas DIG attains a steady state in only 1000 steps. Despite being half as coarse as the ESMC grid, the DIG still produces accurate results due to its asymptotic-preserving property. 
Under these conditions, DIG reduces the number of iterations by nearly an order of magnitude compared to ESMC, and the total simulation time is shortened by a factor of 14. 
When Kn is fixed but En increases, the acceleration ratio of DIG over ESMC slightly decreases. This trend can be attributed to the corresponding increases in viscosity~\eqref{mu_kap}, which moderately improves the efficiency of the ESMC method.

% \begin{figure}[t]
%     \centering
%     \includegraphics[width=0.42\textwidth,trim = 100pt 40pt 100pt 40pt,clip]{figures/porous_ESMC_iteration_rho_En0.01_Kn0.5.pdf}
%     \includegraphics[width=0.42\textwidth,trim = 100pt 40pt 100pt 40pt,clip]{figures/porous_DIG_iteration_rho_En0.01_Kn0.5.pdf}
%     \\
%     \includegraphics[width=0.42\textwidth,trim = 100pt 40pt 100pt 40pt,clip]{figures/porous_ESMC_iteration_rho_En0.01_Kn0.05.pdf}
%     \includegraphics[width=0.42\textwidth,trim = 100pt 40pt 100pt 40pt,clip]{figures/porous_DIG_iteration_rho_En0.01_Kn0.05.pdf}
%     \caption{Convergence history of density at $y$=0, when Kn = 0.5 (first row) and Kn = 0.05 (second row) at En = 0.01.}
%     \label{fig: stagnationline_evolution_density}
% \end{figure}

%Figure \ref{fig: stagnationline_evolution_density} illustrates the convergence history of density at $y=0$, obtained from both DIG and ESMC simulations. 

%The magnitude of momentum density becomes larger with the decrease in Kn, En, and the distance from the blocks because the particles are more likely to collide with the blocks and lose speed, leading to entering the adsorption layer of the block. 
%On the other hand, the adsorption layer also becomes more significant with the decreasing Knudsen numbers. A constant En coupled with a decreasing Kn leads to an increase in the reference length, which enlarges the particle-block contact area and consequently traps more particles within the adsorption layer following surface collisions?.

\section{Conclusions and outlook}\label{sec:6}

In summary, we have developed a multiscale scheme DIG to solve the Enskog equation both efficiently and accurately. The overall structure of the stochastic Monte Carlo simulation for the Enskog equation remains unchanged, but it is augmented with a synthetic equation that incorporates higher-order constitutive relations extracted from the Monte Carlo simulation. The deterministic solution of this synthetic equation guides the simulation particles to evolve rapidly toward the steady state. As a result of this two-way stochastic–deterministic coupling, the Monte Carlo simulation not only converges to the steady state much faster but also relaxes the constraint that the spatial cell size must be smaller than the local mean free path. Consequently, the DIG method reduces computational cost by several orders of magnitude in the near-continuum flow regime.

Leveraging the efficiency and accuracy of DIG, we have simulated and analyzed dense gas flow in a simple porous medium, a problem that has rarely been investigated based on the Enskog equation. Our results show that, at large Enskog numbers, the density near solid walls increases significantly, resembling the adsorption layer observed in shale gas extraction and influencing the overall mass flow rate. However, it should be emphasized that in the present study, the adsorption layer arises solely from the excluded-volume effect. To achieve a more realistic description of solid walls, future work may focus on incorporating gas–surface interactions via the Enskog-type collision operator~\cite{Gibelli2008Enskog,Frezzotti2015} or applying density-functional theory for inhomogeneous fluids~\cite{zhaoliGuo2005}. Moreover, the DIG framework can be extended to the Enskog–Vlasov equation to capture non-equilibrium dynamics at liquid–gas interfaces.

% \section*{Acknowledgments}
% acknowledgments

\bibliographystyle{elsarticle-num}
\bibliography{ref}
\end{document}